\documentclass[traditabstract]{aa}
\usepackage{graphicx}
\usepackage{txfonts}
\usepackage{epsfig}
\usepackage{natbib}
\usepackage{color}

\begin{document}

\title{Chemodynamical evolution of the Milky Way disk II: Variations with Galactic radius and height above the disk plane}

\author{I.~Minchev\inst{1}, 
C.~Chiappini\inst{1}, 
M.~Martig\inst{2,3}
}

\institute{Leibniz-Institut f\"{ur} Astrophysik Potsdam (AIP), An der Sternwarte 16, D-14482, Potsdam, Germany
\email{iminchev1@gmail.com}
\and
Centre for Astrophysics \& Supercomputing, Swinburne University of Technology, P.O. Box 218, Hawthorn, VIC 3122, Australia
\and
Max-Planck-Institut f\"{ur} Astronomie, K\"{o}nigstuhl 17, D-69117 Heidelberg, Germany
}

\abstract{
In the first paper of this series (paper I) we presented a new approach for studying the chemodynamical evolution in disk galaxies, focusing on the Milky Way. While in paper I we studied extensively the Solar vicinity, here we extend these results to different distances from the Galactic center, looking for variations of observables that can be related to on-going and future spectroscopic surveys. By separating the effects of kinematic heating and radial migration, we show that migration is much more important, even for the oldest and hottest stellar population. The distributions of stellar birth guiding radii and final guiding radii (signifying contamination from migration and heating, respectively) widen with increasing distance from the Galactic center. As a result, the slope in the age-metallicity relation flattens significantly at Galactic radii larger than solar. We predict that the metallicity distributions of (unbiased) samples at different distances from the galactic center peak at approximately the same value, [Fe/H]~$\approx-0.15$~dex, and have similar metal-poor tails extending to [Fe/H]~$\approx-1.3$~dex. In contrast, the metal-rich tail decreases with increasing radius, thus giving rise to the expected decline of mean metallicity with radius. Similarly, the [Mg/Fe] distribution always peaks at $\approx0.15$~dex, but its low-end tail is lost as radius increases, while the high-end tails off at [Mg/Fe]~$\approx0.45$~dex. The radial metallicity and [Mg/Fe] gradients in our model show significant variations with height above the plane due to changes in the mixture of stellar ages. An inversion in the radial metallicity gradient is found from negative to weakly positive (at $r<10$~kpc), and from positive to negative for the [Mg/Fe] gradient, with increasing distance from the disk plane. We relate this to the combined effect of (i) the predominance of young stars close to the disk plane and old stars away from it, (ii) the more concentrated older stellar component, and (iii) the flaring of mono-age disk populations. We also investigate the effect of recycled gas flows on the mean [Fe/H] and find that in the region $4<r<12$~kpc the introduced errors are less than 0.05-0.1~dex, related to the fact that inward and outward flows mostly cancel in that radial range. We show that radial migration cannot compete with the inside-out formation of the disk, exposed by the more centrally concentrated older disk populations, and consistent with recent observations.
}


\titlerunning{Chemodynamical evolution of the Milky Way disk II}
\authorrunning{I. Minchev et al.}

\maketitle

\section{Introduction}
\label{sec:intro}

Galactic archeology aims at understanding the formation and evolution of the Milky Way (MW), where the chemical and kinematical information contained in its stellar component is used as fossil records \citep{freeman02, matteucci12}. Unprecedented amounts of data from a number of ongoing and planned astrometric and spectroscopic Galactic surveys (RAVE - \citealt{steinmetz12}; SEGUE - \citealt{yanny09}; HERMES - \citealt{freeman10}; APOGEE - \citealt{allende08}) will be available soon, especially from Gaia \citep{perryman01} and 4MOST \citep{dejong12}, who together will provide high accuracy 6D kinematics and chemistry for more than $10^7$ disk stars. 

It has been recently recognized that chemical evolution galactic disk modeling must be combined with dynamics. The main reason for this is that numerical simulation have shown that stars do not remain near their birth places, but migrate throughout the disk during their lifetimes. This redistribution of angular momentum has been shown to be caused by the effect of non-axisymmetric disk features, such as spiral structure \citep{sellwood02,roskar08a} and the central bar \citep{mf10, minchev11a, brunetti11, dimatteo13}. This has resulted in efforts to understand how traditional, static, chemical evolution disk modeling couples with dynamics, as discussed in detail in the introduction in the first paper of this series, \cite{mcm13} (hereafter paper I). We also refer the reader to the comprehensive recent reviews by \cite{rix13} and \cite{binney13}.

\subsection{The chemodynamical model of paper I}
\label{sec:paper1}

The chemodynamical model we use in this work was presented in paper I, where we mostly concentrated on an annulus of 2~kpc, centered on the ``solar radius". The main features that make this model unique is the fusion between a state-of-the-art simulation in the cosmological context \citep{martig09,martig12} and a detailed thin-disk chemical evolution model. 

The exact star formation history and chemical enrichment from our chemical model is implemented into the simulation with more than 30 elements assigned to each particle. This novel approach has made it possible to avoid problems with chemical enrichment and star formation rates currently found in fully self-consistent simulations, as described in paper I.

The simulation builds up a galactic disk self-consistently by gas inflow from filaments and mergers and naturally takes into account radial migration processes due to early merger activity and internal disk evolution at low redshift. The last massive merger takes place at $\sim9$~Gyr look-back-time having a disk mass ratio of 1:5. A relatively quiescent phase marks the last 8-9~Gyr of evolution. A number of less violent events (1:70-1:100 mass ratio) are present during this period, with their frequency decreasing with time. The accreted disk component at the final time, estimated at $4<r<16$~kpc, is $\sim3$\% of the total disk mass. In paper I and this work we chemically tag only stars born in situ and do not consider the accreted stellar component, which would introduce an additional complexity in the chemo-kinematic relations. The exploration of the range of parameters arising from considering the accreted component is deferred to a future work.

A central bar develops early on, doubles in size in the last $\sim6$~Gyr, and has a similar length at the final simulation time to that of the MW (see paper I, Fig.~1 rightmost top panel). Snapshots of the disk face-on and edge-on stellar surface density can be seen in Fig.~1 of paper I. An important ingredient, which ensures we capture disk dynamics similar to the MW, is that prior to inserting the chemical evolution model, we rescale the simulation to place the ``solar radius" (8~kpc) just outside the 2:1 outer Lindblad resonance (OLR) of the bar, as is believed to be the case for the MW (e.g., \citealt{dehnen00, mnq07, minchev10}). Consideration of the bar when studying the MW disk is important, since the bar is expected to dominate the disk dynamics within 2-3 scale-lengths through its corotation resonance (CR) and OLR, may drive spiral structure of different multiplicity (e.g., \citealt{masset97,quillen11,minchev12a}), and be responsible for coupling between the vertical and radial motions at preferred locations both in the inner \citep{combes90, quillen14} and the outer disk \citep{minchev12b}.

\subsection{Need for chemo-kinematics information of the entire disk}

\begin{figure*}
\includegraphics[width=18.5cm]{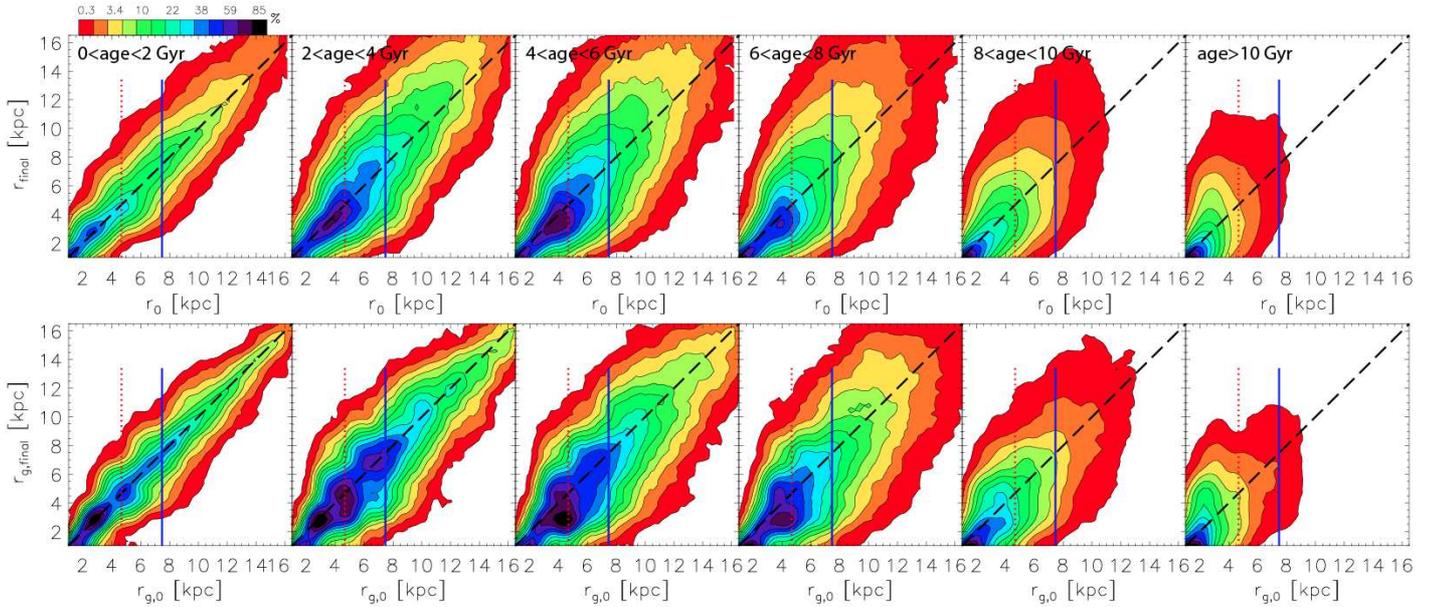}
\caption{
{\bf Top:} Actual birth radii versus final radii for stars in different age bins as indicated in each panel. The dotted-red and solid-blue vertical lines indicate the positions of the bar's CR and OLR. The dashed black line shows the locus of non-migrating circular orbits. {\bf Bottom:} Guiding birth radii versus guiding final radii. Using the guiding radii excludes the effect of high-eccentricity stars, thus exhibiting better the resonant locations as clumps along the dashed black line.
The effect of the bar is seen in the overdensity inside the CR, indicating stars shifted preferentially from their radius outward in the disk. This becomes more evident for younger stars due to their colder orbits, allowing them to respond more strongly to the perturbations. 
}
\label{fig:rirf}      
\end{figure*}

While a range of models may be able to match the chemo-kinematics of stars in the solar neighborhood, discrimination between different evolution scenarios can be made by requiring that models are compliant with data covering extended portions of the disk. By integrating the orbits of RAVE giants, \cite{boeche13b} were able to cover extended disk radii by considering the guiding radii, instead of the current stellar positions. SEGUE G-dwarf data has also been used to cover disk regions between 6-12 kpc \citep{cheng12a, bovy12a}. Large portions of the Galactic disk close to the plane are now observed with APOGEE \citep{anders14, hayden14}.

Variations of chemical gradients are expected for different age populations, as well as for different vertical slices of the disk. A realistic MW chemo-dynamical model must be able to explain not only the local properties of stars but also these variations with Galactic radius and distance from the disk midplane. The goal of this work is to extend the results of paper I to regions beyond the solar vicinity, by using the same model, and provide understanding for the causes of the expected variations. We note, however, that a direct comparison between the results presented here and observational data must be done with care, as observational biases can affect strongly chemo-kinematic relations, especially at large distances. A future work will be dedicated to the proper comparison to observations with the help of mock catalogues constructed from our models (Piffl et al., in preparation).

\section{Effects of radial migration in our simulation}

In paper I (Fig.~1, bottom panel) we showed the changes in angular momentum at different stages of the disk evolution. This revealed that the strength and radial dependence of migration is governed by merger activity at high redshift and internal perturbations from the bar and spirals at later times. 

Another way of looking at the mixing induced throughout the galaxy lifetime is presented in Fig.~\ref{fig:rirf}, where the top row shows density contours of stellar final radii, $r_{final}$, versus birth radii, $r_0$, at the end of the simulation. The contour level separation is given on top of the left row. The inner 1-kpc disk is not shown, in order to display properly the contours. The disk is divided into six age groups, as specified on top of each panel. The dotted-red and solid-blue vertical lines indicate the positions of the bar's CR and OLR at the end of the simulation. The dashed-black line shows the locus of non-migrating circular orbits. Deviations from this line are caused not only from stars migrating from their birth places, but also from stars on eccentric orbits away from their guiding radii. The latter effect becomes more important with increasing age, especially for stars with age~$>8$~Gyr, e.g., those formed before the last massive merger event in our simulation.
 
To exclude the effect of high-eccentricity stars, in the bottom row of Fig.~\ref{fig:rirf} we show the density of final versus birth {\it guiding radii}, $r_{g,final}$ and $r_{g,0}$, respectively. We estimate these for each stellar particle as
\begin{equation}\label{rg0}
r_{g,0} = \frac{L_0}{v_{c,0}}
\end{equation}
and
\begin{equation}\label{rg1}
r_{g} = \frac{L}{v_{c}},
\end{equation}
where $L_0\equiv r_0 v_{\phi,0}$ and $L\equiv r v_{\phi}$ are the initial and final angular momenta, respectively, while $v_{c,0}$ and $v_c$ are the initial and final values of the circular velocity at the corresponding radii, which changes roughly between 190 and 220 km/s, while remaining significantly flat except for the decline at $r<1.5$~kpc (see top row of Fig.~1 in paper I).

Several interesting differences are seen between the stellar distributions in the $r_0-r_{final}$ and $r_{g,0}-r_{g,final}$ planes. Firstly we find that all guiding radius distributions are much more symmetric around the dashed-black line, indicating more balanced exchange of particles across a given disk radius. This is because the asymmetric drift effect, where outer radii are populated by high-eccentricity stars close to their apocenters, has been removed by using the guiding radii. Another difference is the better identification of efficient migration radii in the guiding radii plots (such as the bar's CR and OLR radii indicated by the dotted-red and solid-blue verticals). Note that these become more clear for younger (and thus colder) populations, because colder orbits respond more strongly to the non-axisymmetric perturbations (e.g., spirals and bar). Interestingly, a well defined peak is found inside the bar's CR for the oldest stellar component when we use the guiding radii. While some of these stars have gained angular momentum already during the strong merger phase ending in the first 2-3 Gyr of evolution, the peak becomes well defined only after the bar is formed, as seen in middle row, left panel of Fig.~15 in paper I. This indicates that the bar affects not only the stars born after its formation, but also the oldest stars, as should be expected.

In both rows of Fig.~\ref{fig:rirf} we find that the largest width around the black-dashed line occurs for stars in the age groups $4<age<6$ and $6<age<8$~Gyr. This is indicative of large changes in angular momentum related to significantly long exposure to migration mechanisms. The two oldest populations do not migrate as much because of their high velocity dispersions resulting from the high-redshift merger phase.

A decrease in disk scale-length is apparent with increasing age, ending up with a very concentrated disk for ages~$>$10~Gyr. This is in agreement with observations in the MW (e.g., \citealt{bensby11, bovy12a}) and will be discussed more in Sec.~\ref{sec:hd}. The effect of the bar is seen in the overdensity inside the CR (dotted-red vertical), indicating stars shifted preferentially from their birth radii outward in the disk.

\subsection{Migration cools the disk during mergers}
\label{sec:mer}

We showed in paper I that stars born in the inner disk can arrive at the simulated solar vicinity with lower velocity dispersion than the in-situ born population. This is in drastic contrast to the expected effect of outward migrators in a quiescent disk evolution, where stars arriving from the inner disk are slightly warmer than the non-migrating population \citep{minchev12b}. Below we investigate this in grater detail studying the entire disk.

\begin{figure}[h!]
\includegraphics[width=8.5cm]{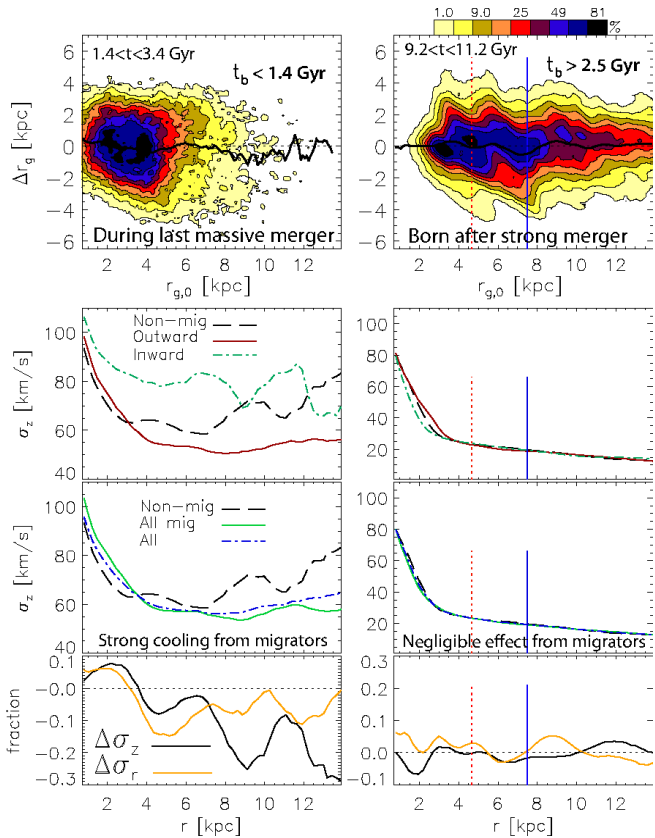}
\caption{
The effect of migration during (left) and after (right) a massive merger. {\bf Left column:} The top panel shows the changes in angular momentum in a time period of 2~Gyr encompassing the merger event. The vertical velocity dispersion profiles of inward migrators, outward migrators and the non-migrating population, as indicated, are shown in the second panel. The net effect of migrators can be seen in the third panel. The bottom panel shows the fractional change in velocity dispersion resulting from migration $\Delta\sigma_{\rm z}=(\sigma_{\rm z,all}-\sigma_{\rm z,non\_mig})/\sigma_{\rm z,all}$. Migrators cool the outer disk, thus working against disk flaring. {\bf Right column:} Same as on the left, but for stars born after the last massive merger event. Minimal effect from migration is seen on the disk vertical velocity dispersion.  
\label{fig:rdsigz}
}
\end{figure}

First we consider all stars born right before the last massive merger encounters the disk, at $t=1.4$~Gyr. To see how much angular momentum redistribution takes place during this merger event, in the top panel of Fig.~\ref{fig:rdsigz} we plot number density contours of the changes in guiding radius, $\Delta r_g$, versus the initial guiding radius, $r_{g,0}$, estimated in the time period $1.4<t<3.4$~Gyr. The percentage of stars in each contour level is given by the color bar on the right. 

The strong redistribution of angular momentum seen in the $r_{g,0}-\Delta r_g$ plane is caused both by the tidal effect of the satellite, which plunges through the galactic center, and the strong spiral structure induced in the gaseous component. We note that the disk dynamics is dominated by the very massive gas disk at this early stage of disk evolution.

Next we separate migrating from non-migrating stars (in the considered time period) by applying the technique described by \cite{minchev12b}. This consists of separating stars in a given radial bin into "migrators" and "non-migrators" as follows. Non-migrators are those particles found in the selected radial bin at both the initial and final times, while migrators are those that were not present in the bin initially but are there at the final time. We distinguish between "outward" and "inward" migrators -- those initially found at radii smaller or larger than the annulus considered, respectively. This is done for radial bins over sampling the entire radial extent of the disk. See \cite{minchev12b} for more details.

\begin{figure*}
\includegraphics[width=18cm]{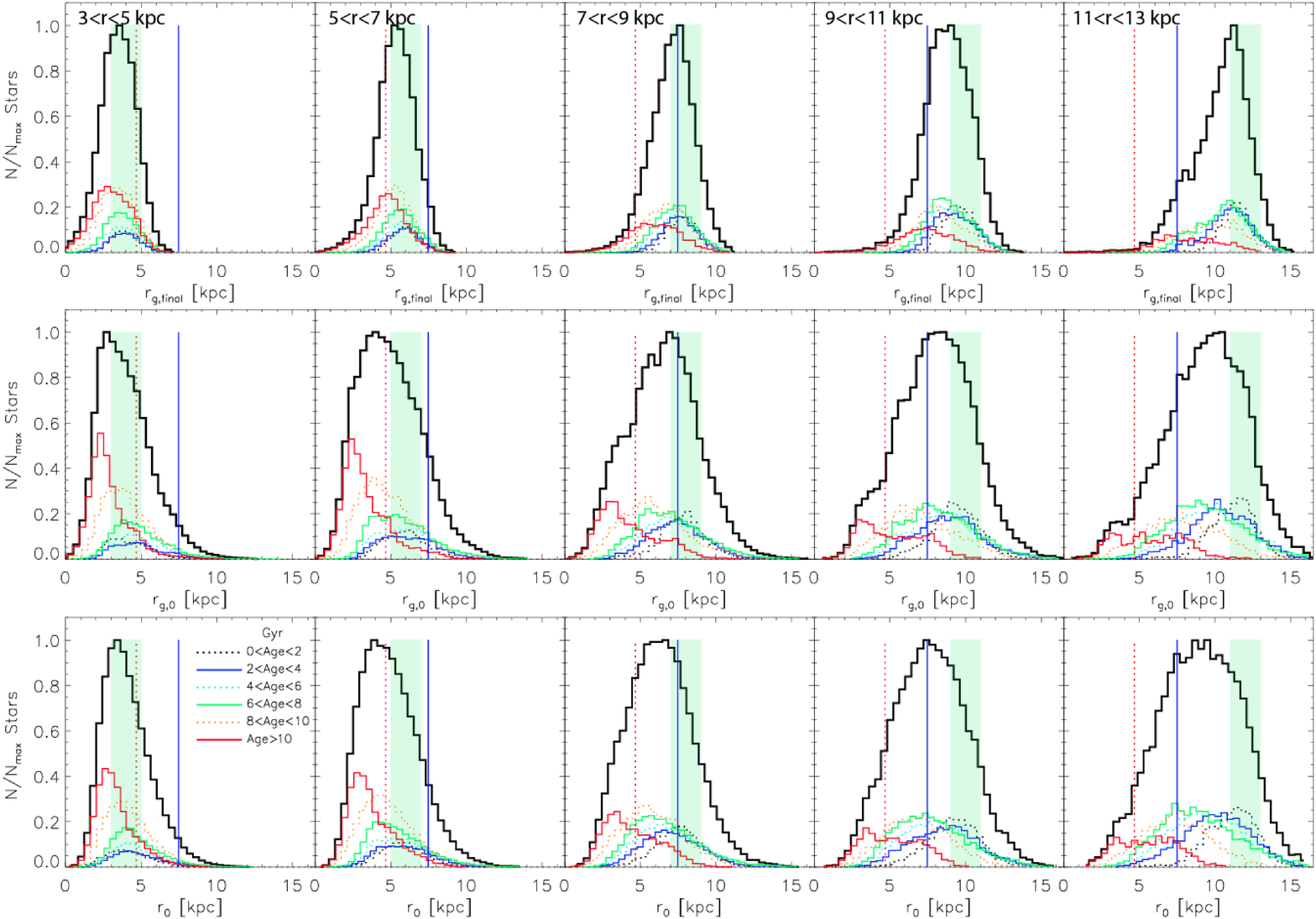}
\caption{
Comparison among the distributions of final guiding radii ($r_{g,final}$, top row), birth guiding radii ($r_{g,0}$, middle row), and actual birth radii ($r_{0}$, bottom row) for stars of different age groups ending up in a given radial bin, as indicated. Five final radial bins are considered. Solid black curves plot the total distribution, while the color-coded curves show the distributions of stars in five different age groups. The dotted-red and solid-blue vertical lines indicate the positions of the bar's CR and OLR estimated at the final time. The width and peak shift of the total distributions are shown in Fig.~\ref{fig:r0a}. The significantly narrower distributions of final guiding radii (compared to $r_{g,0}$) in all age-groups indicate that migration is important for stars of all ages. 
}
\label{fig:r0}      
\end{figure*}

In the second top-to-bottom left panel of Fig.~\ref{fig:rdsigz} we plot the vertical velocity dispersion profiles of inward migrators, outward migrators and the non-migrating population, as indicated. It is interesting to see that inward and outward migrators {\it during the merger} have reversed roles, where stars migrating inwards have positive contribution to $\sigma_z$ and those migrating outward cool the disk. This is the opposite of what is expected for migration in the absence of external perturbations (as described in the Introduction and shown in paper I). The recent work by \cite{minchev14a} showed that the fact that cool, old, [$\alpha$/Fe]-enhanced, metal-poor stars can arrive from the inner disk can provide a way to quantify the MW merger history, where the idea was applied to a high-quality selection of RAVE giants \citep{boeche13a} and the SEGUE G-dwarf sample \citep{lee11}. Indeed, the effect of satellites perturbing the MW disk has been linked to a number of observations of structure in the phase-space of local stars (e.g., \citealt{minchev09, gomez12a,gomez12b,gomez13,widrow12,ramya12}). 

The net effect of migrators during merger can be seen in the third top-to-bottom left panel of Fig.~\ref{fig:rdsigz}. The overall contribution to the vertical velocity dispersion from the migrating stars during the merger is negative, in the sense that it is lower than that of the stars which did not migrate. We emphasize that we only considered stars born before the merger took pace, therefore, the effect seen is not related to the accreted population. 

Finally, to quantify the changes to the disk vertical velocity dispersion resulting from migration in the given period of time, we plot the fractional changes in the bottom left panel of Fig.~\ref{fig:rdsigz}. We estimate these as 
\begin{equation}
\label{eq:1}
\Delta\sigma_{\rm z}=(\sigma_{\rm z,all}-\sigma_{\rm z,non\_mig})/\sigma_{\rm z,all}, 
\end{equation}
where $\sigma_{\rm z,all}$ and $\sigma_{\rm z,non\_mig}$ are the vertical velocity dispersions for the total population and the non-migrators, respectively. Decrease in $\sigma_z$ of up to 30\% can be seen.

We have found here that during a massive merger sinking deep into the disk center, migrators cool the outer disk, thus working against disk flaring. This is related to the stronger effect of mergers on the outer disk, owing to the exponential decrease in the disk surface density. Stars arriving from the inner parts during a merger, would therefore be cooler than the rest of the population.

We now contrast the above results with a sample born after the last massive merger even. The right column of Fig.~\ref{fig:rdsigz} is the same as the left one, but for stars born after $t=2.5$~Gyr (or 8.7 Gyr ago). Minimal effect from migration is seen on the disk vertical velocity dispersion during the last 2~Gyr of (mostly quiescent) evolution, despite the strong migration efficiency seen in the top right panel.

In accordance with the above results, small (and mostly negative) contribution of radial migration on the age-velocity relation was found also by \cite{martig14b}, where a suite of seven simulation similar to the one considered in this work was studied. Additionally, studying the high-resolution isolated disk simulation of \cite{donghia13}, the recent work by \cite{vera-ciro14} also found minimal effect of migration on the disk thickening.

\begin{figure}
\includegraphics[width=8.6cm]{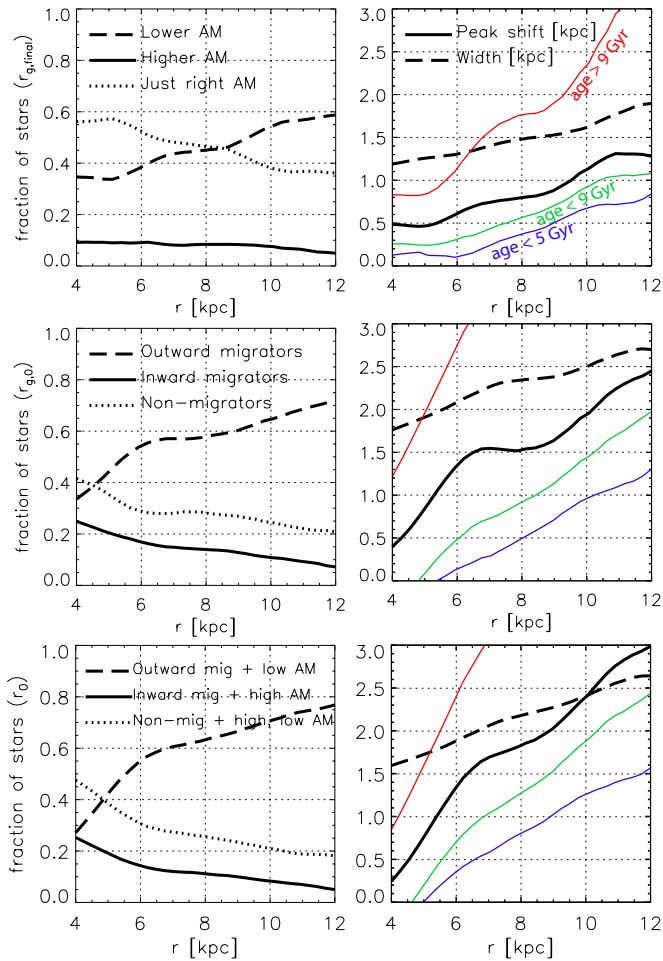}
\caption{
{\bf Left column:} Variation with Galactic radius of the fraction of stars arriving from the inner disk, from the outer disk, and those native to a given radius. {\bf Right column:} Peak shift and width of the total distributions (black curves). The peak shift of stars born before (red) and after (blue) the last massive mergers are also shown. From top to bottom, different rows show the above quantities estimated for the final guiding radius ($r_{g,final}$), the guiding birth radius ($r_{g,0}$), and the actual birth radius ($r_{0}$), as in Fig.~\ref{fig:r0}. The peak shift is defined as the difference between the median final radius and the peak of the total distributions shown in Fig.~\ref{fig:r0}. The width is estimated as the standard deviation of the total distribution in each bin. For this figure we consider 30 overlapping radial bins instead of the five used in Fig.~\ref{fig:r0}, in order to better exhibit the radial variations. 
}
\label{fig:r0a}      
\end{figure}

\subsection{Contamination from migration at different distances from the Galactic center}

We would now like to quantify the radial migration as a function of galactic radius in our simulation. For this purpose, in  Fig.~\ref{fig:r0} we study the origin of stars ending up in five radial bins of width 2 kpc, centered on $r=3$, 5, 7, 9, and 11 kpc. For each annulus (green vertical strip), six different age-groups are shown, color-coded in the bottom-leftmost panel. The solid-black line shows the total distribution in each panel. The bar's CR and OLR are shown by the dotted-red and solid-blue verticals. The middle row corresponds to the solar vicinity (the bottom-middle panel is the same as the left panel of Fig.~3 in paper I) with the bar's OLR at 7.2~kpc. Guiding radii for this figure are estimated from equations \ref{rg0} and \ref{rg1} as in Fig.~\ref{fig:rirf}.

In the top row of Fig.~\ref{fig:r0} we show the {\it final} guiding radii, $r_{g,final}$, of stellar particles ending up in the five annuli considered at the final simulation time. This shows that there is a substantial fraction of stars with angular momentum lower (to the left of the bin) or higher (to the right of the bin) than appropriate for the given annulus, but populating that bin close to their apocenters or pericenters, respectively. The predominancy of stars with lower angular momenta is related to the well-known asymmetric drift effect seen in solar neighborhood stars (mean rotational velocity distribution offset to negative values), which is found here to increase strongly with increasing galactic radius.

In the second row of Fig.~\ref{fig:r0} we show the {\it birth} guiding radii, $r_{g,0}$, of stars currently in the five bins considered. The distributions widen strongly compared to those shown in the first row, indicating significant changes in angular momentum since their birth. The drastic differences found between the distributions in the first and second rows speak of the importance of angular momentum redistribution within the disk, i.e., radial migration. We emphasize the strong changes in angular momentum seen even for the oldest ages (red histograms), although the migration efficiency for these stars is expected to be reduced due to their larger velocity dispersions (see Fig.~\ref{fig:rirf}).

Finally, in the bottom row of Fig.~\ref{fig:r0} we show the actual birth radii, $r_0$, (not guiding radii) of stars ending up in the final five bins considered, where we see the effect of both the changes in angular momentum and heating. Distributions are similar to those of the initial guiding radii, $r_{g,0}$, shown in the second row.

While Fig.~\ref{fig:r0} can be interpreted as giving an account for (1) stars appearing in a given radial bin close to their apo- and pericenters (first row) and for (2) stars which have migrated there (middle row), it must be realized that the situation is more complicated than that. Even stars with guiding radii outside the given radial bin have experienced radial migration. Therefore, those cannot be treated simply as non-migrators contaminating a given radius because of large eccentricities as it is frequently done in the current literature. An evidence that even the oldest and, thus, the hottest stars have migrated substantially is given by the differences seen in the red histogram in the top and middle rows of Fig.~\ref{fig:r0}, where the initial guiding radii cover a much larger radial extent than the final guiding radii.

\begin{figure*}
\includegraphics[width=18cm]{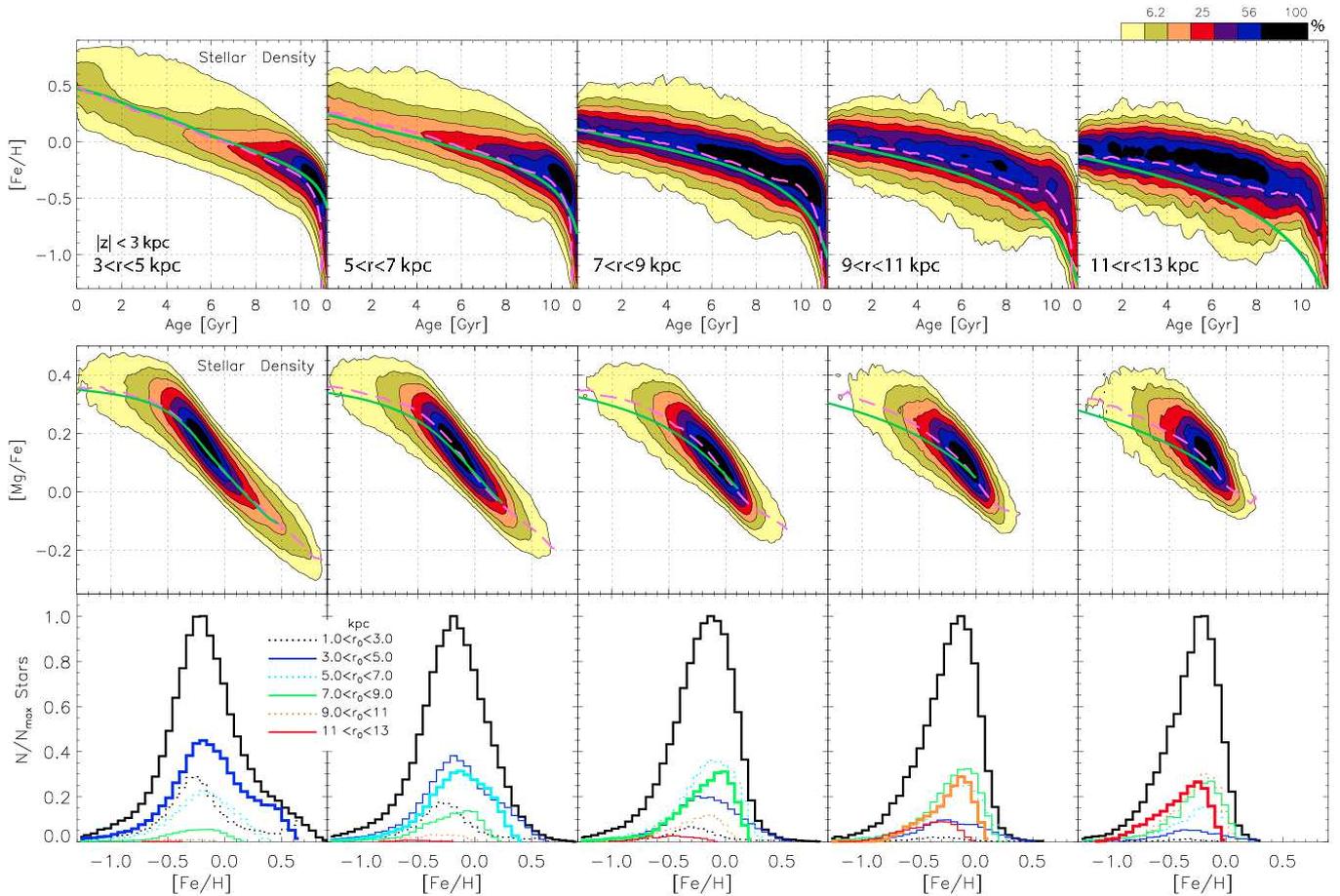}
\caption{
{\bf Top row:} Color contours show the age-[Fe/H] relation for the same disk annuli as in Fig.~\ref{fig:r0}. The middle row ($7<r<9$~kpc) corresponds to the solar neighborhood. The input chemistry, native to each bin, is shown by the solid-green curve. The dashed-pink curve indicates the mean [Fe/H] binned by age. The gradient is only weakly affected at and inward of the solar circle. 
{\bf Second row:} [Fe/H]-[Mg/Fe] stellar distributions. As in the top row, the input chemistry is shown by the solid-green curve. 
{\bf Third row:} Metallicity distributions. In addition to the total sample at each radial bin, the color-coded curves indicate groups born at different galactic radii. In each panel, the thick curve shows the stars born in that given radial bin.
}
\label{fig:chem}      
\end{figure*}

To quantify better the amount of migration and heating seen in Fig.~\ref{fig:r0}, in Fig.~\ref{fig:r0a} we plot the radial variation of the fraction of stars arriving from the inner disk, from the outer disk, and those native to a given radius (left column), as well as the peak shift and width of the distributions (right column). The peak shift is defined as the difference between the median final radius and the peak of the total distribution shown in Fig.~\ref{fig:r0}. The width is estimated as the standard deviation of the total distribution at each annulus.

As in Fig.~\ref{fig:r0}, the three rows of Fig.~\ref{fig:r0a}, from top to bottom, correspond to $r_{g,final}$, $r_{g,0}$, and $r_0$. To see better the radial variation of the above quantities, we have considered 30 overlapping radial bins in the radial range $4<r<12$~kpc, instead of the five bins shown in Fig.~\ref{fig:r0}.

Focussing on the top-left panel of Fig.~\ref{fig:r0a}, we find an increase with increasing distance from the Galactic center, in the fraction of stars with {\it final guiding radii} ($r_{g,final}$) inside a given annulus (red-dashed curve), from $\sim$35\% to $\sim$60\%. Only about 10\% of stars have guiding radii higher than those appropriate for a given annulus (blue-solid curve). The fraction of stars with angular momenta just right for a given annulus decreases with Galactic radius from $\sim$55\% to $\sim$35\% (black-dotted curve). Note that at the solar radius (8 kpc) the fraction of stars belonging to the inner disk is similar to that of stars with angular momenta typical for that annulus. It should be kept in mind that the above numbers will change if we were to use an annulus width different than the 2-kpc utilized for Figures~\ref{fig:r0} and \ref{fig:r0a}. 
 
While we find that a lot of contribution can come from stars with angular momenta lower than that appropriate for a given radial bin (e.g., $\sim50\%$ at the solar radius), it is also important to know how far these stars come from. Therefore, in the top-right panel of Fig.~\ref{fig:r0a} we show the radial variation of the distributions' peak shift (solid curve) and width (dashed curve). An increase in both of these is found with radius, indicating stronger contamination in the outer disk.

In the second row of Fig.~\ref{fig:r0a} we show that the fractions of stars with {\it birth guiding radii} ($r_{g,0}$) coming from the inner or outer disk are appreciably larger than the fraction of stars arriving at a given radial bin simply due to their eccentric orbits (as in the top row). For example, at the solar radius about 60\% of stars were born in the inner disk, only about 30\% were born in situ ($7<r<9$~kpc), and about 15\% migrated from the outer disk. However, in the second-row, right panel, we see that the peak shift at the solar radius more than doubles and the width of the distribution is also quite larger than that seen in the final guiding radii. The nonlinear increase with radius of the peak shift is related to the variation in migration efficiency with Galactic radius, as seen in the bottom row of Fig.~1 of paper I (see discussion in paper I). The rate of increase in contamination with radius is higher for the birth guiding radii than for the final guiding radii by about a factor of two (compare slopes in black-solid curves in the top and middle panels in the right column of Fig.~\ref{fig:r0a}). 

Finally, the bottom row of Fig.~\ref{fig:r0a} shows the fraction of stellar actual birth radii ($r_{0}$) contributing to different radial bins. This can be seen as the effect of both migration and heating, but note that it is not simply the addition of the top two rows since, for example, some stars with guiding radii appropriate for a given radial bin can be found outside that annulus at a given time. The bottom row of Fig.~\ref{fig:r0a} is similar to the middle row, with somewhat higher fraction of stars arriving from the inner disk (red-dashed curve in left panel) and yet steeper increase in the peak shift (black-solid curve in right panel). 

As mentioned earlier and seen in Fig.~\ref{fig:r0}, the oldest stars are the ones most affected by radial migration processes. To illustrate this and contrast the effect of migration and heating on young and old stars, in the right column of Fig.~\ref{fig:r0a} we over-plotted the peak shift variation with radius of stars born before the last massive merger (ages$>9$~Gyr, solid-red curve), stars born after the last massive merger (ages$<9$~Gyr, solid-green curve), and stars with age$<5$~Gyr (solid-blue curve). It is notable that (i) the increase of the peak shift with radius for the younger stellar populations is also quite strong and (ii) the effect is significantly stronger when migration and heating act together -- a maximum of $\sim1.6$~kpc shift is found for $r_0$ (bottom panel), while the effect of heating only, indicated by $r_{g,final}$, reaches $\sim$0.85. 

The predicted increase in contamination from migration and heating with radius is due to the exponential drop in disk surface density, where more stars are available to migrate outwards than inwards. Note that inward migration is still very important both for the kinematics and chemistry, in that stars arriving from the outer disk balance the effect of stars coming from the inner disk at intermediate radii (as discussed in Sections~\ref{sec:amr} and \ref{sec:gas}). 

As may be expected at this point, this difference in contamination from radial migration and heating at different galactic radii should have strong effect on the disk final chemistry. We anticipate these effects to be now testable with the large body of data coming from current Galactic spectroscopic surveys. We show the implication of this in the following sections.

\section{Radial variations of chemodynamical relations}

\subsection{Age-Metallicity Relation at different radii}
\label{sec:amr}

The first row of Fig.~\ref{fig:chem} plots age-[Fe/H] stellar density contours for the same disk annuli as in Fig.~\ref{fig:r0}. Contour levels are indicated in the color-bar attached to the rightmost panel. The middle panel ($7<r<9$~kpc) corresponds to the solar neighborhood and is the same as the plot shown in Fig.~4 of paper I. The input chemistry, native to each bin, is shown by the green-solid curve. The pink dashed curves indicate the mean [Fe/H] in each panel. Measurement uncertainties of $\pm0.1$ dex drawn from a uniform distribution are convolved with our simulated [Fe/H].

While a scatter in the age-metallicity relation (AMR) is seen at all radii, the mean (pink-dashed curves) is very close to the local evolution curve in the inner three bins. Only outside the solar radius do we find significant deviation, with the strongest flattening in the AMR at the outermost bin ($11<r<13$~kpc). This results from the accumulation of outward migrators in the outer disk, as seen in Figures~\ref{fig:r0} and \ref{fig:r0a}. Conversely, at radii close to and smaller than the solar radius the contribution from metal-rich stars arriving from the inner disk is mostly balanced by metal-poor stars arriving from the outer disk. This is analogous to the contribution to the velocity dispersion from inward and outward migrators discussed by \cite{minchev12b}, where the overall effect on the disk heating (and thus thickening) was found to be minimal at radii less than about four disk scale-lengths. Similar effect results for the gas (as discussed in Sec.~\ref{sec:gas}, see Fig.~\ref{fig:gas}), rendering recycled gas flows unimportant in the range $4<r<12$~kpc.   

\begin{figure}
\includegraphics[width=8.6cm]{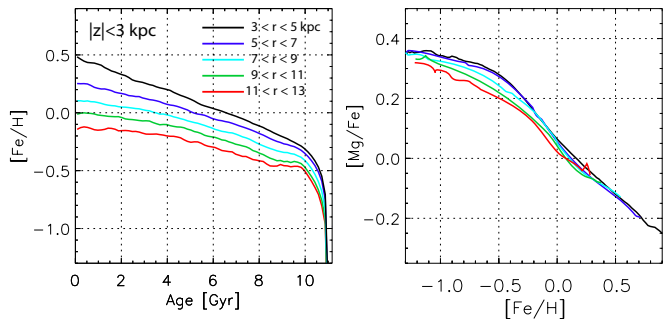}
\caption{
{\bf Left:} The mean [Fe/H] variation with age at different Galactic radii as color-coded. Identical to the pink-dashed curves in the top row of Fig.~\ref{fig:chem}. {\bf Right:} The mean [Fe/H]-[Mg/Fe] relation for different radii. Same as pink-dashed curves in the secind row of Fig.~\ref{fig:chem}.
}
\label{fig:chem1a}      
\end{figure}

The left panel of Fig.~\ref{fig:chem1a} shows the mean [Fe/H] variation with age at different Galactic radii as color-coded. These curves are identical to the pink-dashed curves in the top row of Fig.~\ref{fig:chem}.

Another important property of the AMR predicted by our model is that the maximum metallicity achieved in each bin decreases from [Fe/H]$\sim$0.5-0.6 to about $\sim$0.2, as one moves from the innermost towards the outermost radial bins. The exact amount of metal rich stars predicted by our model is sensitive to the initial chemical model assigned to the inner regions of the disk. As discussed in paper I, we extrapolated our thin-disk chemical evolution model to the Galactic center in the innermost 2~kpc, not assigning a specific bulge chemistry to particles born in that region. This is here justified by the fact that we first want to study the effect of migration in a pure disk. The impact of considering bulge chemistry will be studied in a forthcoming paper. We anticipate that this would mostly affect the predicted fraction of super-metal-rich stars in the radial bins internal to the solar vicinity.

The fraction of stars with metallicities above $\sim$0.2-0.3~dex can be used as a powerful constraint of our models. Observationally, it is still difficult to quantify this fraction not only due to observational biases induced by color-selected spectroscopic samples, but also because of difficulties of current abundance pipelines to account for metallicities above solar. This situation will certainly improve in the future as spectral libraries and stellar isochrones get extended beyond solar.

\begin{figure}
\includegraphics[width=8.5cm]{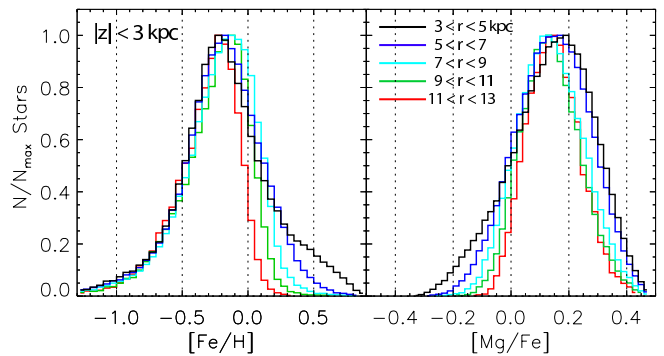}
\caption{
{\bf Left:} Overlaid normalized metallicity distributions at different distances from the Galactic center (all solid-black histograms in bottom row of Fig.~\ref{fig:chem}). The peak of the total distribution for each radial bin is always centered on [Fe/H]$\approx-0.15\pm0.06$~dex and the metal-poor wings are nearly identical, extending to [Fe/H]$\approx-1.3$~dex. In contrast, the metal-rich tail of the distribution decreases with increasing radius, giving rise to the expected decrease in mean metallicity with radius. {\bf Right:} As on the left but for [Mg/Fe]. Similarly to [Fe/H], the peak does not vary significantly as a function of Galactic radius, being situated at [Mg/Fe]$\approx0.15\pm0.08$~dex. Unlike the metallicity distribution, the $\alpha$-poor tail is lost as radius increases, while the $\alpha$-rich tail always ends at [Mg/Fe]$\approx 0.45$~dex. However, the correspondence between [Fe/H] and [Mg/Fe] is not symmetric: for [Mg/Fe] both wings of the distribution are affected and the width decreases with increasing radius. See Fig.~\ref{fig:chem4} for variation with distance from the disk plane.
}
\label{fig:chem3}      
\end{figure}

\begin{figure*}
\includegraphics[width=18cm]{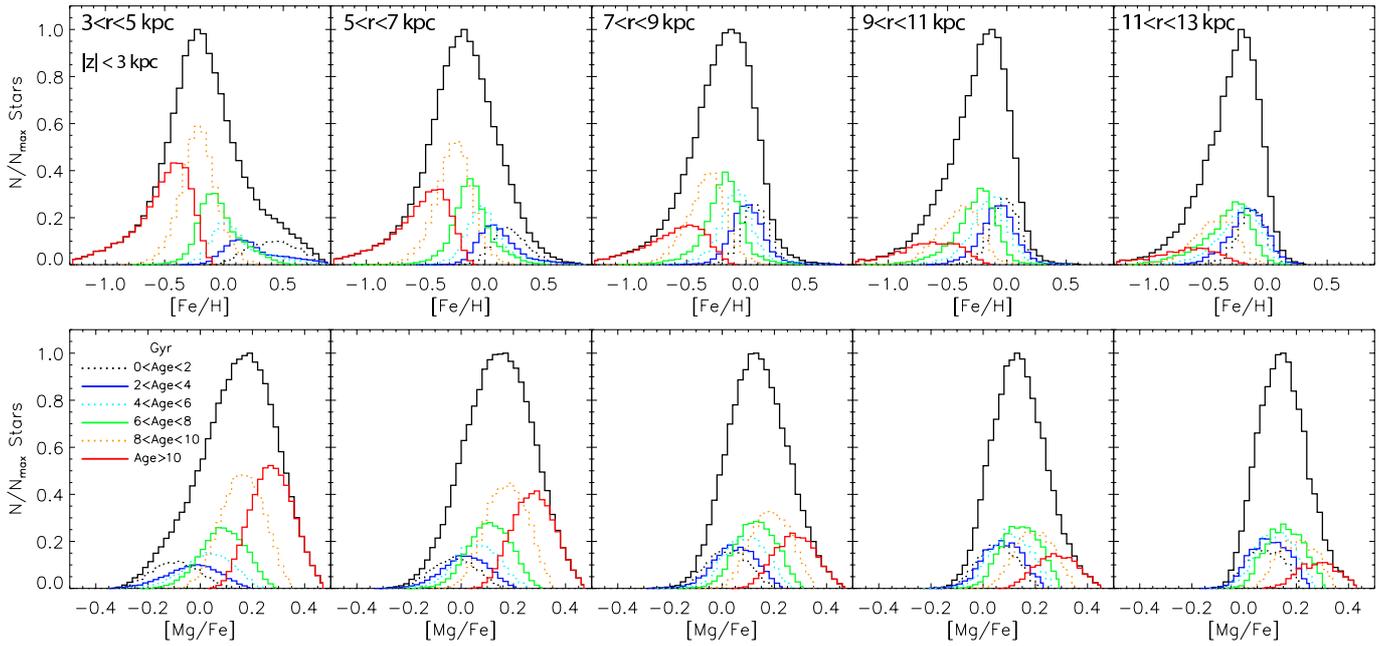}
\caption{
{\bf Top:} Metallicity distributions at different distances from the Galactic center. In each panel the solid black histogram shows the total sample (same as colored lines in Fig.~\ref{fig:chem3}, left), while various colors correspond to different age groups as indicated in the bottom left panel. It is remarkable that despite the strong decrease in the fraction of old stars (red and orange lines), the metal-poor tail remains practically identical for samples at different radii (as seen in Fig.~\ref{fig:chem3}, left). While in the central regions it is dominated by the oldest stars, in the outer disk it is a mixture of all ages. {\bf Bottom:} Same as above but for [Mg/Fe]. Here the decrease in the fraction of old stars gives a significant effect in the $\alpha$-rich wing of the distribution (see Fig.~\ref{fig:chem3}, right).
}
\label{fig:chem2}      
\end{figure*}

\subsection{[Fe/H]-[Mg/Fe] relation at different radii}

The second row of Fig.~\ref{fig:chem} shows [Fe/H]-[Mg/Fe] density contours. Uncertainties of $\pm$0.1 and $\pm$0.05 dex (typical of high-resolution observations in the literature) drawn from a uniform distribution are convolved with our simulated [Fe/H] and [Mg/H]  abundances, respectively\footnote{These uncertainties are the same as those used for [Fe/H] and [O/H] in paper I, although there it was erroneously stated that $\pm0.05$~dex was the uncertainty in [O/Fe]}.

An overall contraction in [Mg/Fe] is found with increasing galactocentric distance, due to the decrease of high-metallicity stars in the outer bins as well as the decrease of high-[Mg/Fe] stars.

In the model, most of the super-metal-rich stars originating in the innermost disk regions (see gold contours) are predicted to have sub-solar [Mg/Fe] ratios. The exact amount of this effect (i.e., the absolute values of [Mg/Fe] at a different [Fe/H]) is strongly dependent on the adopted stellar yields of Mg, Fe and the SNIa rates (here we adopt the same stellar yields as in \citealt{francois04}). How stellar yields behave at above solar metallicity regimes is still very uncertain. Improvements on stellar evolution models at very high-metallicities are on the way by several groups, and will be soon implemented in our models as well.

As a summary of the second row of Fig.~\ref{fig:chem}, the left panel of Fig.~\ref{fig:chem1a} shows the variation of the mean [Fe/H]-[Mg/Fe] relation with Galactic radius as color-coded. These curves are identical to the pink-dashed curves in the second row of Fig.~\ref{fig:chem}. Only small variations are found, mostly at super-solar values of [Mg/Fe].

We do not find a bimodality in the [Fe/H]-[Mg/Fe] at any distance from the Galactic center. We showed in paper I that the distribution at the solar radius can become bimodal when selection criteria used in high-resolution surveys (e.g., \citealt{bensby03}) were applied (see paper I, Fig.~12). It should be noted that the gap in [$\alpha$/Fe] (usually $\sim0.2$~dex) has now been seen in a number of different observational samples (e.g., SEGUE - \citealt{lee11}, APOGEE - \citealt{anders14}, HARPS - \citealt{adibekyan13}, GES - \citealt{recio-blanco14}) indicating that it may be a real feature and not just a selection bias.

Several studies of the [Fe/H]-[$\alpha$/Fe] plane (e.g., \citealt{bensby11,cheng12b,anders14}) have shown that the number of high-[$\alpha$/Fe] metal-poor stars decreases strongly in the outer disk. A work in preparation is dedicated to the proper comparison with observations, where survey biases will be considered when comparing to our model. However, a decline in the maximum value of [Mg/Fe] is already clearly seen in the second row of Fig.~\ref{fig:chem} - a shift downward in [Mg/Fe] of $\sim0.2$~dex is found in the three densest contours, when comparing the innermost to the outermost radial bins. Introducing a gap at $\sim0.2$~dex (as discussed above), which would result naturally from a gap in the star formation at high redshift (as in the two-infall model of \citealt{chiappini97}), may naturally improve the comparison between our model and the observations.

\subsection{[Fe/H] and [Mg/Fe] distributions at different radii}

The third row of Fig.~\ref{fig:chem} shows the metallicity distributions at different distances from the Galactic center. In each panel the solid-black histogram shows the total sample, while various colors correspond to groups of common birth radii, as indicated on the left. The thick histogram in each panel shows stars born in that given radial bin, e.g, green corresponds to the solar vicinity.
 
In all final radial bins the metal-rich tail of the distribution results from stars originating in the inner disk (compare the local metal-rich tail to that of the total distribution). Note that the largest contribution in the range $7<r<11$~kpc (the solar bin and the neighboring two bins) to the metal-rich tail comes from the bar CR region (blue curve). Therefore, the existence of a metal-rich tail throughout most of the disk can be linked to the effect of the MW bar. This was already noted in paper I for the solar vicinity (see bottom row of Fig.~1 of that work).

\begin{figure}
\includegraphics[width=8.5cm]{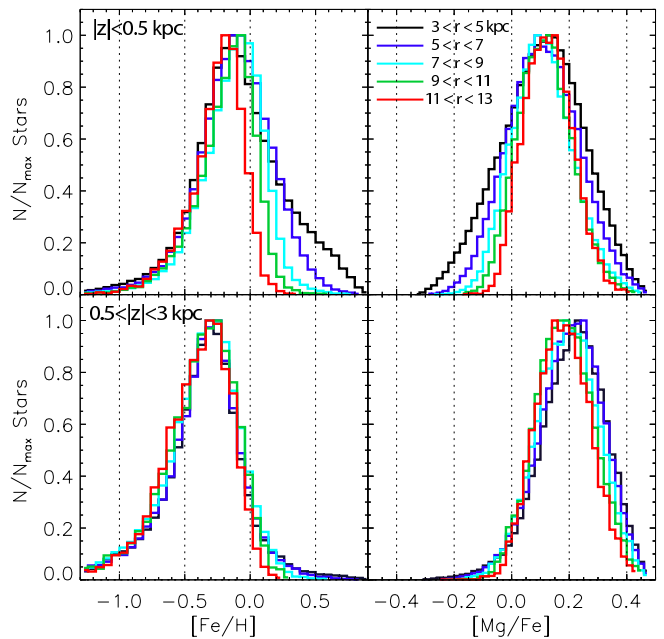}
\caption{
{\bf Left column:} Overlaid normalized metallicity distributions for different distances from the Galactic center  for stars with $|z|<0.5$~kpc (top) and $0.5<|z|<3$~kpc (bottom). This shows that variations in the metal-rich end seen in Fig.~\ref{fig:chem3} come mostly from stars close to the disk midplane. 
{\bf Right column:} As on the left but for [Mg/Fe]. Similarly to [Fe/H], strong variations with radius are seen mostly for stars close to the disk midplane. The reason for this is that migrating stars stay with cool kinematics (both in the radial and vertical directions), i.e., do not contribute to thick disk formation.
}
\label{fig:chem4}      
\end{figure}

For better comparison of the [Fe/H] distributions at different distances from the Galactic center (solid black histograms in the bottom row of Fig.~\ref{fig:chem}), in the left panel of Fig.~\ref{fig:chem3} we show these overlaid and color-coded, as indicated in the right panel. We find that the peak of the total distribution for each radial bin is always centered on [Fe/H]$\approx-0.15\pm0.06$~dex and the metal-poor wings are nearly identical, extending to [Fe/H]$\approx-1.3$~dex. In contrast, the metal-rich tail of the distribution decreases with increasing radius, giving rise to the expected decrease in mean metallicity with radius. This is a property of the chemical evolution model used, as evident from examining the locally born stars in each annulus in the bottom panel of Fig.~\ref{fig:chem}. However, we can also see clearly that for each annulus the metal-rich tail is extended (in each case by about 0.2-0.3 dex) because of stars migrating from the inner disk. As shown later in Fig.~\ref{fig:chem4}, this results from stars close to the disk midplane - another evidence that outward migrators do not populate a thick disk. 

The right panel of Fig.~\ref{fig:chem3} is like the left one but for [Mg/Fe]. Similarly to [Fe/H], the peak does not vary significantly as a function of Galactic radius, being situated at [Mg/Fe]$\approx0.15\pm0.08$~dex. In this case the $\alpha$-poor tail is lost as radius increases, while the $\alpha$-rich tail always ends at [Mg/Fe]$\approx 0.45$~dex. However, the correspondence between [Fe/H] and [Mg/Fe] is not symmetric: for [Mg/Fe] both wings of the distribution are affected. In both cases the distributions get broader with increasing radius, which is a testable prediction with current Galactic spectroscopic surveys.

To see what ages comprise different regions of the [Fe/H] and [Mg/Fe] distributions, in Fig.~\ref{fig:chem2} we decompose them into six different age groups, as indicated in the bottom left panel. It is remarkable that despite the strong decrease in the fraction of old stars (red and orange lines) with increasing radius, the metal-poor tail remains practically identical for samples at different radii (as seen in Fig.~\ref{fig:chem3}, left). While in the central regions it is dominated by the oldest stars, in the outer disk it is a mixture of all ages. 

In the bottom row of Fig.~\ref{fig:chem2} we see that the decrease in the fraction of old stars with increasing radius gives a significant effect in the $\alpha$-rich wing of the distribution (see Fig.~\ref{fig:chem3}, right). On the other hand, the $\alpha$-poor tail becomes more prominent in the inner radial bins due to the contribution of the youngest stars (with ages below 3-4 Gyr, born inside the solar circle; see also Fig.~\ref{fig:chem2}).

The strong decrease of the fraction of old/[Mg/Fe]-rich stars with Galactic radius seen in Fig.~\ref{fig:chem2} indicates that in our model the age/chemically defined thick disk has a shorter scale-length than the thin disk. This is in agreement with observations in the Milky Way (e.g., \citealt{bensby11, anders14, bovy12a}) and will be discussed further in Sections \ref{sec:hd} and \ref{sec:highmg}.

\begin{figure*}
\includegraphics[width=18cm]{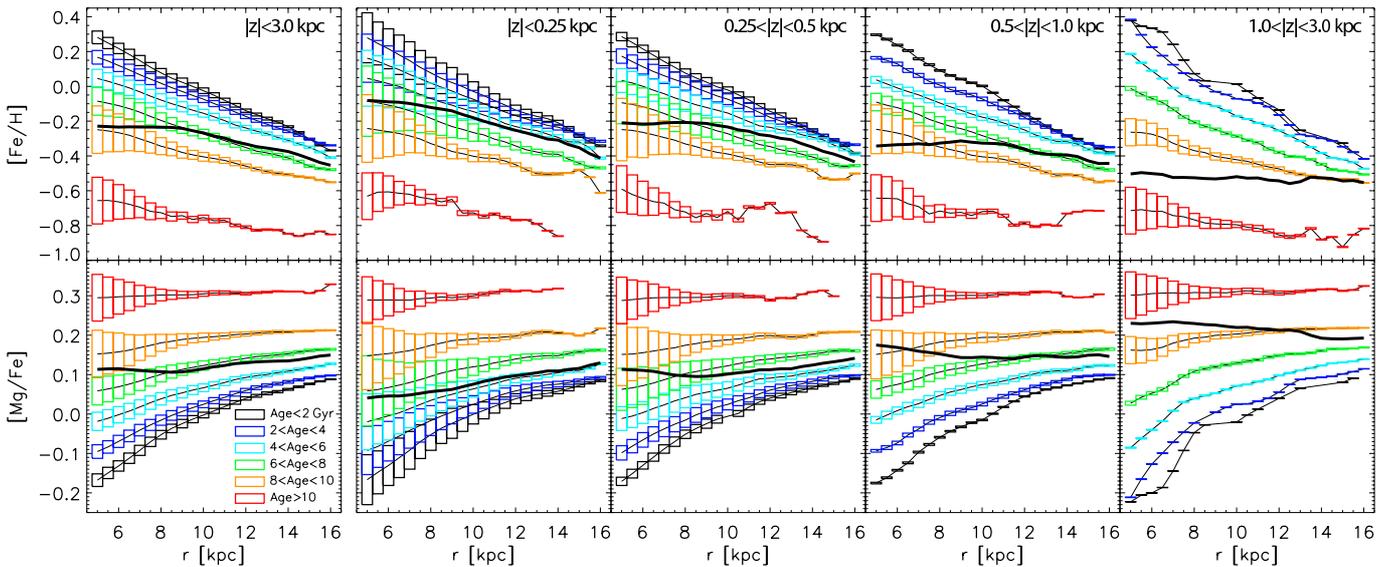}
\caption{
Variation of azimuthally averaged chemical gradients with distance from the disk midplane. {\bf Top row:} Thick black curves show the metallicity variation with galactic radius for stellar samples at different distances from the disk midplane, as marked in each panel. Different colors correspond to different age groups as indicated in the bottom-left panel. The height of rectangular symbols reflects the density of each bin. The negative gradient seen in the total population (leftmost panel) is strongly flattened (and even reversed at $r<10$~kpc) for the range $0.5<|z|<1.0$~kpc. {\bf Bottom row:} Same as above but for [Mg/Fe]. Here the weak positive gradient for the total population (leftmost panel) turns negative as distance from the disk plane increases, although the gradient of each individual age-bin is positive. This figure illustrates that different mixtures of stellar ages (e.g., because of different slices in $z$) can give rise to a range of different [Fe/H] and [$\alpha$/Fe]   gradients.
}
\label{fig:grad}      
\end{figure*}

\section{Variations with distance from the disk midplane}

In the previous sections we showed general chemodynamical properties of our Galactic disk model by analyzing all particles confined to 3~kpc vertical distance from the plane (where most of the disk particles lie and current surveys cover). In this section we consider different cuts in the vertical direction.

\subsection{[Fe/H] and [Mg/Fe] distributions at different heights}

We showed in Fig.~\ref{fig:chem3} how the [Fe/H] and [Mg/Fe] distributions vary with distance from the Galactic center. Because the high-[Fe/H] and low-[Mg/Fe] tails come from stars migrating from the inner disk, it is interesting to find out how this is reflected in samples at different distances from the disk plane.

Similarly to the left panel of Fig.~\ref{fig:chem3}, in the left column of Fig.~\ref{fig:chem4} we show normalized metallicity distributions for different distances from the Galactic center, but for stars with $|z|<0.5$~kpc (top) and $0.5<|z|<3$~kpc (bottom). We find that variations in the high-[Fe/H] tail seen in Fig.~\ref{fig:chem3} come mostly from stars close to the disk midplane. The right column of Fig.~\ref{fig:chem4} is the same as the left one, but for [Mg/Fe]. Similarly to [Fe/H], strong variations with radius are seen mostly for stars close to the disk midplane. The reason for this is the fact that stars close to the disk are not just vertically cool, but also their eccentricities are low. Stars with low eccentricities are also the youngest (on the average), and thus the most metal-rich population (on the average). In contrast, the older/hotter population reaching high distances away from the plane is metal-poor. Note that if stars heated the outer disk as they migrate outwards (i.e., if they populated regions high above the disk midplane) there would not be a difference in the samples close and away from the disk plane, and thus no negative vertical metallicity gradient as seen in a number of observations (e.g., \citealt{carraro98,chen11,kordopatis11}).

The age-[Fe/H] and [Fe/H]-[Mg/Fe] relations, as well as the corresponding distributions for stars with $|z|<0.5$ and $0.5<|z|<3$~kpc are shown in Figures~\ref{fig:chemz1} and \ref{fig:chemz2} in appendix~\ref{sec:a1}. 

\subsection{Chemical gradients at different heights above the plane}

Various studies have found different metallicity and [$\alpha$/Fe] gradients in the MW, as discussed in the Introduction. Fig.~\ref{fig:grad} illustrates that different mixtures of stellar ages (e.g., because of different slices in $z$) can give rise to a range of different [Fe/H] and [$\alpha$/Fe] gradients.

Thick black curves in the top row show the azimuthally averaged metallicity variation with galactic radius for stellar samples at different distances from the disk midplane, as marked in each panel. Different colors correspond to different age groups as indicated in the bottom-left panel. The height of rectangular symbols reflects the density of each bin. The bottom row of Fig.~\ref{fig:grad} shows the same information as above but for [Mg/Fe].

\begin{table*}
\centering
\small
\vspace{0.5cm}
\footnotesize
\label{tab:fr}
\begin{tabular}{c c c c c c}
\hline
  &  $|z|<3$~kpc &   $|z|<0.25$~kpc  &  $0.25<|z|<0.5$~kpc & $0.5<|z|<1.0$~kpc & $1.0<|z|<3.0$~kpc \\
\hline
        &  [Fe/H], [Mg/Fe] & [Fe/H], [Mg/Fe] & [Fe/H], [Mg/Fe] & [Fe/H], [Mg/Fe] & [Fe/H], [Mg/Fe]\\
\hline
${\rm age}<2$~Gyr &  $-0.058, 0.028$  &  $-0.057, 0.027$  &  $-0.059, 0.028$  &   $-0.064, 0.030$ & $-0.077, 0.038$ \\
$2<{\rm age}<4$    &  $-0.048, 0.021$ &  $-0.047, 0.021$  &  $-0.048, 0.021$   &   $-0.049, 0.021$ & $-0.070, 0.033$\\
$4<{\rm age}<6$    &  $-0.040, 0.015$ &  $-0.040, 0.015$  &  $-0.038, 0.014$   &   $-0.041, 0.015$ & $-0.058, 0.023$\\
$6<{\rm age}<8$    &  $-0.039, 0.012$  &  $-0.038, 0.012$  &  $-0.037, 0.011$   &   $-0.038, 0.011$ & $-0.050, 0.016$\\ 
$8<{\rm age}<10$  &  $-0.031, 0.007$  &  $-0.032, 0.008$  &  $-0.028, 0.007$   &   $-0.030, 0.007$ & $-0.032, 0.007$\\
${\rm age}>10$      &  $-0.022, 0.002$   &  $-0.025, 0.004$  &  $-0.012, 0.001$   &   $-0.020, 0.002$ & $-0.020, 0.001$\\
\hline
All ages    & $-0.016, 0.003 $ & $-0.027, 0.009$ & $-0.012, 0.001 $ & $-0.004, -0.004$ & $-0.006, -0.003$\\
\hline
\end{tabular}
\caption{
Gradients in [Fe/H] and [Mg/Fe] (dex kpc$^{-1}$) for different age populations and distances from the disk midplane, corresponding to Fig.~\ref{fig:grad}. Radial range of $5<r<13$~kpc was used for fitting. For a given age subsample, only minimal variations are seen with distance from the disk. However, gradients in the total population can vary strongly due to (i) the predominance of young stars close to the disk plane and old stars away from it (as illustrated by the rectangular symbols in Fig.~\ref{fig:grad}) and (ii) the more concentrated older stellar component. It should be kept in mind that the variations of [Fe/H] and [Mg/Fe] with radius are rarely well fitted by a single line in both the observations and our model (see Fig.~\ref{fig:grad}). Therefore, non-negligible variations in the estimated gradients should be expected with a change in the radial range used for fitting.   
}
\end{table*}

The negative radial metallicity gradient seen in the total population (leftmost upper panel) is strongly flattened with increasing $|z|$, and even reversed at $r<10$~kpc for the range $0.5<|z|<1.0$~kpc. By examining the variation in density of different age subsamples, we see that the change in slope with increasing $|z|$ is caused by the strong decrease of stars with ages~$<6$~Gyr at $|z|>0.5$~kpc for $r\lesssim12$~kpc. 

Focusing on [Mg/Fe] we find that the weak positive gradient for the total population (leftmost bottom panel in Fig.~\ref{fig:grad}) turns negative as distance from the disk plane increases, although the gradient of each individual age-bin is positive.

In contrast to the strong variation of chemical gradients with distance from the disk midplane for stars of all ages, the gradients of individual age groups do not vary significantly with distance from the plane (see Table~1). Gradients in the total population can vary strongly with $|z|$ because of the interplay between (i) the predominance of young stars close to the disk plane and old stars away from it (as illustrated by the rectangular symbols in Fig.~\ref{fig:grad}), (ii) the more concentrated older stellar component (as seen in Fig.~\ref{fig:den} below), and (iii) the flaring of mono-age disk populations (see \citealt{martig14a}, Fig.~5). 

\begin{figure*}
\includegraphics[width=14cm]{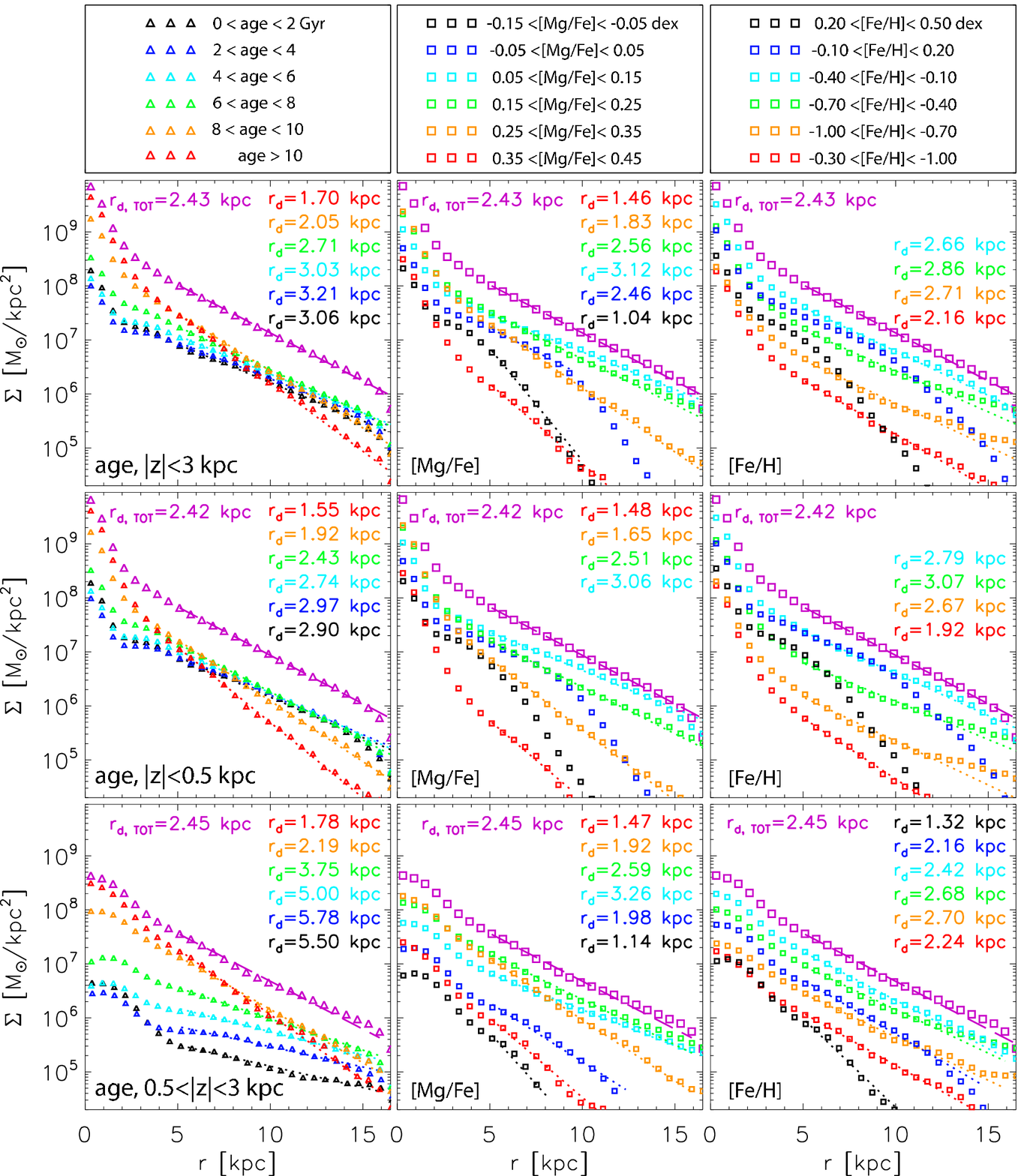}
\centering
\caption{
{\bf First row:} Stellar surface density as a function of galactic radius for stars with distance from the midplane $|z|<3$~kpc. In addition to the total mass (pink symbols), subsamples grouped by narrow bins of age (left panel), [Mg/Fe] (middle panel), and [Fe/H] (right panel) are shown by different colors, as indicated. The corresponding bin values are shown on top of the figure. The color-coded values in each panel indicate the scale-length, $r_d$, fitted as a single exponential in the range $5<r<16.5$~kpc, shown by the dotted lines (dashed pink line for the total population). Decrease in scale-length is seen with increasing age for all three distances from the disk plane.
While single exponential fits are appropriate for mono-age populations (left column), stars binned by [Mg/Fe] or [Fe/H] show truncation (a change to a steeper exponential) around 5-10~kpc for the younger stars (black and blue curves).
{\bf Second row:} Same as above, but for stars with $|z|<0.5$~kpc. {\bf Third row:} Same as above, but for stars with $0.5<|z|<3$~kpc. At this distance from the disk plane (similar to SEGUE G-dwarfs), the breaks in the high-[Fe/H] and low-[Mg/Fe] profiles are mostly gone.
}
\label{fig:den}      
\end{figure*}

The above discussion suggests that the results of different Galactic surveys should be compared with care, taking into account the spatial and magnitude coverage of observational samples. Indeed, before large spectroscopic surveys were in place, most of the MW abundance gradients reported in the literature were obtained by using rather young populations, such as Cepheids (e.g., \citealt{andrievsky02, pedicelli09, luck11, lemasle13}), HII regions  (e.g., \citealt{daflon04, stasinska12} and references therein) and the more numerous young open clusters (e.g., \citealt{jacobson11, yong12, frinchaboy13}). The advantage of these young population tracers is that they cover a large radial disk extent and are mostly located in the disk midplane, i.e., lie in a very low $z$-range. It is clear from Fig.~\ref{fig:grad} that such young tracers will yield steeper metallicity gradients than a mixed population. 

The situation changes when using field stars of mixed ages. In addition the the very local GCS, populations of mixed ages were covered in larger regions around the solar neighborhood thanks to RAVE and SEGUE. Using data from the latter two surveys, it became possible to infer abundance gradients of field stars at high distances from the plane (although most of the region near the Galactic plane, i.e., $|z|< $~0.2-0.4 kpc is not sampled in this case, see \citealt{cheng12a, boeche13b, boeche14}). Recently, abundance gradients for field stars (all ages covering vertical distances from zero to beyond 3~kpc are being estimated thanks to the SDSS-APOGEE survey \citep{hayden14, anders14} -- an infrared high-resolution survey that can observe stellar populations very close to the galactic plane, filling the gap left by SEGUE and RAVE. Note that in GCS, RAVE, SEGUE and APOGEE, the age mix is not only a strong function of the distance from the plane and from the galactic center, but also dependent on the selection biases of each sample.

Given the above discussion, we do not show in the present paper a direct comparison of the magnitude of our predicted gradients with observations. The main focus here is to understand what is driving the general shape of the abundance gradients given the different mix of ages of the tracer populations at different heights from the plane. However, we can say that the [Fe/H] gradients predicted for our youngest population bin ($<$ 2 Gyr), in the range $5<r<13$ kpc (see Table 1) are in good agreement with the values reported in the literature for Cepheids (e.g., \citealt{lemasle13}) and young open clusters \citep{frinchaboy13} of around $-0.06$~dex/kpc. For [Mg/Fe] we predict a positive gradient of $\sim0.03$~dex/kpc for stars with age~$<2$~Gyr, which could be in slight tension with the observations of young populations showing almost flat [Mg/Fe] gradients (e.g., \citealt{jacobson11}). Our values for ''all ages" also compare well with the recent values reported in the literature by \cite{boeche13b} and \cite{anders14}, based on RAVE and APOGEE data, respectively, both for iron and [$\alpha$/Fe] gradients. A more detailed comparison with RAVE and APOGEE data is deferred to a future work, where we will take properly into account the spatial and magnitude coverages (along with the expected sample biases).

It should also be kept in mind that the variations of [Fe/H] and [Mg/Fe] with radius are rarely well fitted by a single line in both observations and our model. Therefore, non-negligible variations in the estimated gradients should be expected with a change in the radial range used for fitting.  

\begin{figure*}[ht]
\includegraphics[width=17.cm]{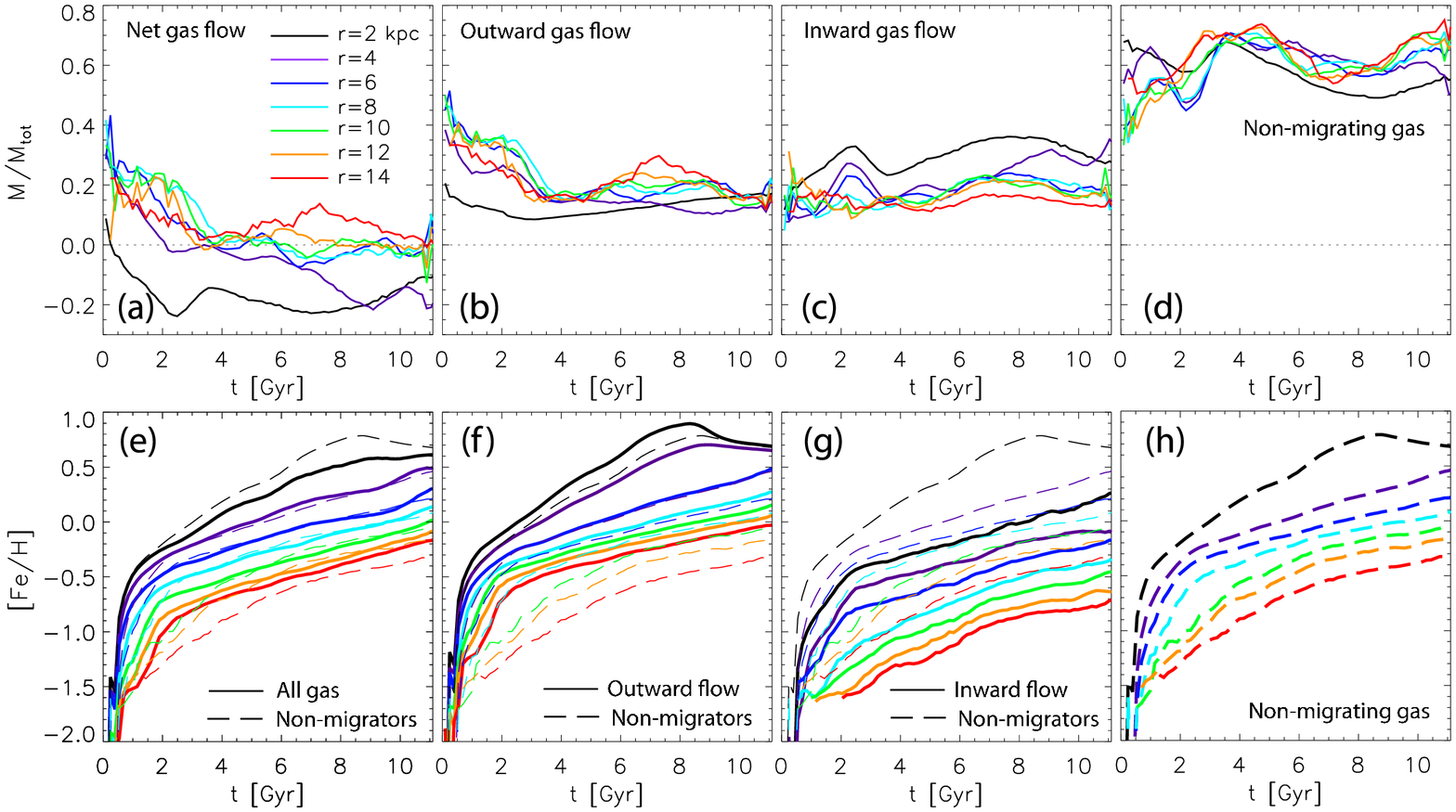}
\centering
\caption{
The effect of recycled gas flows on the input [Fe/H] used in our model:
{\bf Top row:} 
Fraction of net gas flow (panel a) into 7 disk annuli of width 2~kpc as a function of time (panel a), where negative values correspond to net inflows. Also shown are the fractions of gas migrating into each bin from inside (i.e., outward migrators, panel b), from outside (i.e., inward migrators, panel c), and gas which does not leave the bins (panel d).
{\bf Bottom row:} 
Mean metallicity estimated from the total gas mass (panel e), the gas arriving from inside the given bin (panel f), from outside the given bin (panel g), and from the gas which stays in the bin (panel h). The overlaid dashed curves in panels (e), (f), and (g) show the non-migrating gas (same as panel h). 
Strong deviations from the input chemistry is found for outward and inward flows (by more than 0.5 dex). However, the net effect is drastically reduced as seen in panel (e), where in the range $4<r<12$~kpc (purple to orange curves) the deviations from the input curves is less than $0.1$~dex during most of the time evolution. 
}
\label{fig:gas}      
\end{figure*}

\subsection{Disk scale-lengths at different heights above the plane}
\label{sec:hd}

We now examine the variation of disk scale-length with distance from the disk plane for populations grouped by common ages or chemistry. 

The first row of Fig.~\ref{fig:den} shows stellar surface density as a function of galactic radius for stars with distance from the midplane $|z|<3$~kpc. In addition to the total mass (pink symbols), stars grouped by six bins of age (left panel), [Mg/Fe] (middle panel), and [Fe/H] (right panel) are shown by different colors. The corresponding bin values are indicated above each panel. The second and third rows of Fig.~\ref{fig:den} are the same as the first one, but for stars with $|z|<0.5$ and $0.5<|z|<3$~kpc, respectively. The color-coded values in the left column, $r_d$, indicate single exponential fits in the range $5<r<16.5$~kpc, shown by the dotted lines (dashed pink line for the total population). 

The total density has $r_d=2.43$~kpc at $|z|<3$~kpc, however, depending on the age, values can range from $\sim1.7$ to $\sim3.1$~kpc. This smooth increase in scale-length with decreasing age is a manifestation of the disk inside-out growth and is in agreement with the results of \cite{bovy12a}, if we assume that mono-age populations correspond to mono-abundance populations. Note that despite the significant migration throughout the disk evolution (see Figures~\ref{fig:rirf} - \ref{fig:r0}), older disks do not increase scale-lengths fast enough to compete with the naturally resulting larger scale-lengths for younger populations, due to the changes in the SFR as a function of radius. Therefore, while migration {\it does flatten} surface density profiles (e.g, \citealt{foyle08, debattista06, minchev12a}), this cannot overtake the effect of inside-out disk growth. Deviations from this rule for some intermediate-age populations was reported by \cite{martig14a}, possibly related to satellite-disk interactions.

Decrease in scale-length is seen with increasing age for all three distances from the disk plane (Fig.~\ref{fig:den}, left column). The youngest population shows the largest scale-length at $0.5<|z|<3$~kpc ($r_d=5.5$~kpc) and the smallest at $|z|<0.5$ ($r_d=2.9$~kpc), which indicates that the fraction of young stars at larger radii increased with distances from the disk plane. This suggests that coeval younger populations should flare, e.g., increase their scale-height with radius. Indeed, this was shown by \cite{martig14a} for the same simulation we study here. 

When stars are binned by [Mg/Fe] (top row, middle panel), we see a break in the trends we have found so far. While for narrow bins of [Mg/Fe] in the range 0.45-0.05 single exponentials can still be fit well at $r>5$~kpc, this is no longer true for lower [Mg/Fe] values. The lowest two bins (blue and black squares) deviate from the increasing trend in scale-length for samples with decreasing [Mg/Fe]. A downtrend at $\sim10$ and $\sim5$~kpc is found for the blue and black squares, respectively. For the sake of comparison, for these two bins we still fit single exponentials in the range $5<r<10$~kpc. 

Similar difficulties with fitting single exponentials to the highest [Fe/H] bins are also found in the right column of Fig.~\ref{fig:den}. The situation is very similar for stars with $|z|>0.5$~kpc (Fig.~\ref{fig:den}, bottom middle-row panel). 

As we move away from the disk plane (as in the SEGUE G-dwarf sample, for example), we find that both [Mg/Fe] and [Fe/H] can all be fit reasonably well by single exponentials. However, the correspondence between age and chemistry is not anymore valid for high [Fe/H], low [Mg/Fe] stars, which do not show large scale-lengths as the youngest stellar age groups.

\subsection{Gas flows}
\label{sec:gas}

We now investigate how radial gas flows may affect our results. In our simulation stellar mass loss is considered by converting stellar particles to gas throughout the simulation (see \citealt{martig12}). This can be used to estimate the neglect of gas flows in our chemodynamical model, by studying the effect of the radial motion of this "enriched" gas. We considered the gas converted from stars at each time step and assigned chemistry to it as a function of time and disk radius, similarly to what we did for the stars in paper I.

Panel (a) in the top row of Fig.~\ref{fig:gas} shows the fraction of net gas flow per unit of time, into 7 disk annuli of width 2~kpc as a function of time. We estimate this as the difference between the mass of gas coming from the inner disk and gas coming from the outer disk, therefore negative values correspond to net inflows. Also shown in the top row are the fractions of gas migrating into each bin from inside (i.e., outward migrators, panel b), from outside (i.e., inward migrators, panel c), and the gas which does not leave the bins (panel d).

Typically, about 60-70\% of gas in each annulus does not migrate and contributions from the inner and outer disks are about 20-30\%, with generally slightly larger fraction of gas migrating outward. Exceptions to this rule are found for the innermost and outermost annuli.  
Consequently, the net flows into the annuli considered are close to zero (panel a), except for the innermost and outermost disk regions. Larger fraction of outward migrating gas is seen during the first couple of Gyr of disk formation, i.e, during the merger epoch.

In the bottom row of Fig.~\ref{fig:gas} we now show the mean metallicity estimated from the total gas mass (panel e), the gas arriving from inside the given bin (panel f), from outside the given bin (panel g), and from the gas which stays in the bin (panel h). The overlaid dashed curves in panels (e), (f), and (g) show the non-migrating gas (same as panel h). 

As expected, strong deviations from the input metallicity are found for outward only, or inward only flows (by more than 0.5 dex). However, the net effect is drastically reduced as seen in panel (e), where in the range $4<r<12$~kpc (purple to orange curves) the deviations from the input curves is less than $0.1$~dex during most of the time evolution. This is caused by the fact that the effects of inward and outward flows mostly cancel in the radial range $4<r<12$~kpc. 
 
From the above discussion we conclude that our results for intermediate radial distance from the Galactic center (including for the solar neighborhood presented in paper I) are not affected to any significant level by the neglect of gas flows. However, gas flows may be affecting our results for the innermost and outermost radial bins we study in this paper. 

We note that inflows of pristine (metal-poor) gas from cosmic web filaments may contribute at the disk outskirts by decreasing the metallicity there, as also reasoned by \cite{kubryk13}. This can counteract the effect of the radial-migration-induced gas flows to the disk outer boundary seen in Fig.~\ref{fig:gas}. On the other hand, increase in density in the inner 2 kpc due to inward migration of both gas and stars can be expected to increase the SFH in that region. This in turn can result in higher chemical enrichment (higher metallicity) counteracting the flattening of metallicity gradients induced by migration (see \citealt{cavichia13}). It can be argued, therefore, that the neglect of gas flows does not have a strong impact on our results. Further work is needed to investigate the above ideas.

\section{Fraction and velocity dispersions of old/high-[Mg/Fe] stars at different radius and distance from the disk midplane}
\label{sec:highmg}

We here consider the stars in our model with thick-disk-like chemistry. Similarly to \cite{fuhrmann11} (see his Fig.~15), we used a definition for the thick disk [$\alpha$/Fe]$<0.2$~dex (here we use [Mg/Fe]) and considered the volume defined by $7.9<r<8.1$~kpc, $|z|<0.1$~kpc. This resulted in a local fraction of thick to total disk mass of 14\%, which is somewhat lower than the value of 20\% estimated by \cite{fuhrmann11} for his volume-complete local sample. As the distance from the solar position in our model increases from 0.1~kpc to 3~kpc, the thick disk mass fraction increases from 14\% to 27\%. These numbers are shown in Table~2.

We next divided the disk into inner ($4<r<8$~kpc) and outer parts ($8<r<15$~kpc), and considered four different maximum distances from the disk midplane, $|z|$. The thick to total disk mass fractions for different $(r,|z|)$ ranges are displayed in Table~2, where it can be seen that (1) close to the midplane there is hardly any variation, (2) as $|z|$ increases the fraction of high-[Mg/Fe] stars always increases, and (3) at $|z|>0.5$~kpc the inner disk contains more stars with thick-disk-like chemistry compared to the outer disk. The overall decrease of high-[Mg/Fe] stars in the outer disk is consistent with the results of recent observational studies (e.g., \citealt{bensby11, anders14, bovy12a}), which have shown that the chemically defined Milky Way thick disk is more centrally concentrated than the thin disk. In our model this results naturally due to the smaller disk scale-lengths of older populations (and given the correspondence between age and [Mg/Fe]), which we discussed in Sec.~\ref{sec:hd}.

To see the difference between a chemistry- and an age-defined thick disk, in Table 2 we also list the fraction of stars with age$>9$~Gyr with respect to the total disk mass. We pick this age cut because it marks the last massive merger as discussed in paper I and in Sec.~\ref{sec:paper1}. While at the solar radius separation by [Mg/Fe] and age gives virtually the same thick to total disk mass fraction as a function of $|z|$, in the inner disk the age definition results in $\sim30\%$ more massive thick disk and in the outer disk in about 20\% less massive. This is to say that in our model the age-defined thick disk is more centrally concentrated than the chemically defined one. In both cases the thin disk is more extended.

We also estimated the vertical velocity dispersions of the chemistry- and age-defined thin and thick disks in each one of spatial regions defined above. These numbers are also given in Table~2. Increase in velocity dispersion is seen with decreasing Galactic radius, as expected. However, only small variations are found with vertical distance from the disk midplane (for a given radius), which is consistent with expectation for the Milky Way \citep{bovy12c}.

While here we used a purely chemical or age definition of the thick disk, it must be kept in mind that the situation is more complicated than that, especially outside the solar neighborhood. It is still unclear how the centrally concentrated chemically/age defined thick disk in the Milky Way reconciles with the extended kinematically defined thick disks in observations of external galaxies. 
 
\begin{table*}
\centering
\small
\vspace{0.5cm}
\footnotesize
\label{tab:fr}
\begin{tabular}{c c c c}
\hline
  &   Inner disk & Near the solar radius & Outer disk \\
  &  $4<r<8$~kpc      & $7.9<r<8.1$~kpc        & $8<r<15$~kpc \\
  \hline
   && Chemistry-defined thick disk: fr$_{\rm [Mg/Fe]>0.2}$, $\sigma_{\rm z,thin}$~[km/s], $\sigma_{\rm z,thck}$~[km/s]  &\\
   \hline
   $|z|<0.1$~kpc  &    0.14, 25, 53   &  0.14, 20, 36  &  0.13, 16, 33  \\
   $|z|<0.5$~kpc  &    0.20, 30, 54   &  0.16, 22, 43  &  0.16, 17, 34  \\
   $|z|<1.0$~kpc  &    0.26, 32, 56   &  0.22, 23, 42  &  0.19, 19, 35  \\
   $|z|<3.0$~kpc  &    0.30, 35, 58   &  0.27, 25, 48  &  0.25, 20, 40  \\
   \hline
   && Age-defined thick disk: fr$_{\rm age>9}$, $\sigma_{\rm z,thin}$~[km/s], $\sigma_{\rm z,thck}$~[km/s]  &\\
  \hline
  $|z|<0.1$~kpc  &   0.18, 22, 54   &   0.14, 19, 38   &   0.09, 16, 40    \\
  $|z|<0.5$~kpc  &   0.27, 26, 55   &   0.16, 20, 46   &   0.11, 17, 41    \\
  $|z|<1.0$~kpc  &   0.34, 28, 56   &   0.22, 21, 45   &   0.14, 18, 42    \\
  $|z|<3.0$~kpc  &   0.41, 29, 58   &   0.28, 22, 50   &   0.20, 19, 45    \\
\hline
\end{tabular}
\caption{ 
Fraction of thick to total disk mass at different radii and distances from the disk midplane. Two ways of defining the thick disk are used: by chemistry ([$\alpha$/Fe]$<0.2$~dex) and by age (age$>9$~Gyr).
For each spatial region and thick disk definition three values are given: the aforementioned fraction (fr$_{\rm [Mg/Fe]>0.2}$ or fr$_{\rm age>9}$), the vertical velocity dispersion of the thin disk ($\sigma_{\rm z,thin}$), and the vertical velocity dispersion of the thick disk ($\sigma_{\rm z,thick}$).
}
\end{table*}

\section{Conclusions}

In this work we investigated how chemo-kinematic relations change with position in the Galactic disk, by analyzing the chemo-dynamical model first introduced in \cite{mcm13} (paper I). The results of paper I were extended beyond the solar vicinity, considering both variations with Galactic radius and distance above the disk midplane. Our main results are as follows:

$\bullet$
We demonstrated that during mergers stars migrating outwards arrive significantly colder than the in-situ population (Fig.~\ref{fig:rdsigz}), as first suggested in paper I. This also has the important effect of working against disk flaring. Our results that stars migrating outwards in the disk cool the locally born population is in stark contrast to the suggestion that radial migration can form a thick disk. The need for massive mergers at high redshift or stars born hot in a turbulent phase (i.e, during the formation of the thick disk) has been indicated in a number of recent observational studies as well (e.g, \citealt{liu12, kordopatis13a, minchev14a}).

$\bullet$
We investigated the effect of recycled gas flows and found that in the region $4<r<12$~kpc the introduced errors in [Fe/H] are less than 0.05-0.1~dex, related to the fact that inward and outward flows mostly cancel in that radial range. 

$\bullet$
We show that radial migration cannot compete with the inside-out formation of the disk, which can be observed as the more centrally concentrated older disk populations (Fig.~\ref{fig:den}). This is in agreement with recent results from simulations of spiral galaxies (e.g., \citealt{brook12,stinson13,bird13}) and observations in the MW \citep{bensby11, bovy12a}.

$\bullet$ 
Because contamination by radial migration and heating becomes more evident with increasing distance from the galactic center, significant flattening (with respect to the in-situ chemical evolution) in the age-metallicity relation is found outside the solar radius. However, at $r<9$~kpc the slope of the locally evolving population is mostly preserved, due to the opposite contribution of inward and outward migrators.

$\bullet$
We predict that the metallicity distributions of (unbiased) samples at different distances from the Galactic center peak at approximately the same value, [Fe/H]~$\approx-0.15$~dex, and have similar metal-poor tails extending to [Fe/H]~$\approx-1.3$~dex (Fig.~\ref{fig:chem3}). In contrast, the metal-rich tail decreases with increasing radius, thus giving rise to the expected decline of mean metallicity with radius. This effect results predominantly from stars close to the plane (Fig.~\ref{fig:chem4}).

$\bullet$
Similarly to [Fe/H], the [Mg/Fe] distribution always peaks at $\approx0.15$~dex, but its low-end tail is lost as radius increases, while the high-end tails off at [Mg/Fe]~$\approx0.45$~dex. 

$\bullet$
The radial metallicity and [Mg/Fe] gradients in our model show significant variations with height above the plane due to a different mixture of stellar ages (Fig.~\ref{fig:grad}). We find an inversion in the metallicity gradient from negative to weakly positive (at $r<10$~kpc), and from positive to negative for the [Mg/Fe] gradient, with increasing distance from the disk plane. We relate this to the disk inside-out formation, where older stellar populations are more centrally concentrated. This indicates the importance of considering the same spatial region for meaningful comparison between different studies, as well as observations and simulations. 

$\bullet$
In contrast to the strong variation of chemical gradients with distance from the disk midplane for stars of all ages, the gradients of individual age groups do not vary significantly with distance from the plane (see Table~1). Gradients in the total population can vary strongly with $|z|$. We relate this to (i) the predominance of young stars close to the disk plane and old stars away from it, (ii) the more concentrated older stellar component, and (iii) the flaring of mono-age disk populations. 

$\bullet$
The [Fe/H] distributions shift peaks from $\approx-0.15$~dex to $\approx-0.3$~dex when samples move from $|z|<0.5$ to $0.5<|z|<3$~kpc (Fig.~\ref{fig:chem4}). We showed that the strong effect on the high-[Fe/H] tail as a function of Galactic radius comes predominantly from the stars closer to the plane, related to both the chemical model used and the radial migration efficiency.

$\bullet$
Similarly to [Fe/H], the [Mg/Fe] distributions are also affected when varying the sample distance from the disk midplane. When changing $|z|<0.5$ to $0.5<|z|<3$~kpc, the [Mg/Fe] peaks shift from $\approx0.1$~dex to $\approx0.15-0.25$~dex, depending on the sample's distance from the Galactic center.

Finally, we would like to emphasize an important dynamical effect that leads to better understanding of what shapes chemo-kinematic relations.
We showed in paper I and this work, that migration is significant in our simulation for both stars and gas, which in principle should prevent us from recovering the disk chemical history. However, an interesting effect occurs away from the disk's inner and outer boundaries, in the region $4<r<12$~kpc ($\sim$1.5-4 scale-lengths). In this radial range material is exchanged across any given radius in such a way as to approximately preserve the kinematical and chemical properties of the stars and gas that do not migrate, such as stellar velocity dispersions \citep{minchev12b} and mean age-[Fe/H] relations (see Figures~\ref{fig:chem} and \ref{fig:gas}). Evidence of migration can still be found as scatter in the stellar age-[Fe/H] relation or the presence of low-velocity dispersion, $[\alpha$/Fe]-enhanced stars in the solar neighborhood \citep{minchev14a}.

In contrast, at the disk boundaries, stars migrating inwards accumulate in the inner disk (innermost one scale-length) and those migrating outwards accumulate in the disk outskirts (outside 3-4 scale-lengths). These effects should be possible to identify in observations, for example hot populations in disk outskirts if extended disk profiles are caused by radial migration \citep{minchev12b} or flattening in metallicity gradients in the disk outskirts (Sec.~\ref{sec:gas}).

The results presented in this work provide tests for surveys, such as RAVE, SEGUE, HERMES, GES, LAMOST, APOGEE, Gaia, WEAVE, and 4MOST. However, once again, we emphasize that direct comparison between the chemo-kinematic relations presented here and observational data may result in erroneous conclusions. For a proper comparison it is imperative to correct data for observational biases, as done for the SEGUE G-dwarf sample \citep{bovy12a}. Conversely, using the selection function of a given survey, mock observations can be extracted from our models that can be compared directly with the data, which is our choice of tackling this problem (Piffl et al., in preparation). 

\acknowledgements
We thank the anonymous referee for helpful suggestion that have greatly improved the manuscript. We also thank G. Cescutti, B. Gibson, B. Famaey, M. Steinmetz, and R. de Jong for fruitful discussions.


\begin{thebibliography}{}

\bibitem[\protect\astroncite{{Adibekyan} et~al.}{2013}]{adibekyan13}
{Adibekyan}, V.~Z., {Figueira}, P., {Santos}, N.~C., et al.: 2013,
\newblock {\em \aap} {\bf 554}, A44

\bibitem[\protect\astroncite{{Allende Prieto} et~al.}{2008}]{allende08}
{Allende Prieto}, C., {Majewski}, S.~R., {Schiavon}, R., et al.: 2008,
\newblock {\em Astronomische Nachrichten} {\bf 329}, 1018

\bibitem[\protect\astroncite{{Anders} et~al.}{2014}]{anders14}
{Anders}, F., {Chiappini}, C., {Santiago}, B. X., et al.: 2014,
\newblock {\em \aap} {\bf 564}, A115

\bibitem[\protect\astroncite{{Andrievsky} et~al.}{2002}]{andrievsky02}
{Andrievsky}, S.~M., {Kovtyukh}, V.~V., {Luck}, R.~E., {L{\'e}pine}, J.~R.~D.,
  {Maciel}, W.~J., and {Beletsky}, Y.~V.: 2002,
\newblock {\em \aap} {\bf 392}, 491

\bibitem[\protect\astroncite{{Bensby} et~al.}{2011}]{bensby11}
{Bensby}, T., {Alves-Brito}, A., {Oey}, M.~S., {Yong}, D., and {Mel{\'e}ndez},
  J.: 2011,
\newblock {\em \apjl} {\bf 735}, L46

\bibitem[\protect\astroncite{{Bensby} et~al.}{2003}]{bensby03}
{Bensby}, T., {Feltzing}, S., and {Lundstr{\"o}m}, I.: 2003,
\newblock {\em \aap} {\bf 410}, 527

\bibitem[\protect\astroncite{{Binney}}{2013}]{binney13}
{Binney}, J.: 2013,
\newblock {\em \nar} {\bf 57}, 29

\bibitem[\protect\astroncite{{Bird} et~al.}{2013}]{bird13}
{Bird}, J.~C., {Kazantzidis}, S., {Weinberg}, D.~H., {Guedes}, J., {Callegari},
  S., {Mayer}, L., and {Madau}, P.: 2013,
\newblock {\em \apj} {\bf 773}, 43

\bibitem[\protect\astroncite{{Boeche} et~al.}{2013a}]{boeche13a}
{Boeche}, C., {Chiappini}, C., {Minchev}, I., et al.: 2013a,
\newblock {\em \aap} {\bf 553}, A19

\bibitem[\protect\astroncite{{Boeche} et~al.}{2014}]{boeche14}
{Boeche}, C., {Siebert}, A., {Piffl}, T., et al.: 2014,
\newblock {\em ArXiv e-prints}

\bibitem[\protect\astroncite{{Boeche} et~al.}{2013b}]{boeche13b}
{Boeche}, C., {Siebert}, A., {Piffl}, T., et al.: 2013b,
\newblock {\em \aap} {\bf 559}, A59

\bibitem[\protect\astroncite{{Bovy} et~al.}{2012a}]{bovy12b}
{Bovy}, J., {Rix}, H.-W., and {Hogg}, D.~W.: 2012a,
\newblock {\em \apj} {\bf 751}, 131

\bibitem[\protect\astroncite{{Bovy} et~al.}{2012b}]{bovy12a}
{Bovy}, J., {Rix}, H.-W., {Liu}, C., {Hogg}, D.~W., {Beers}, T.~C., and {Lee}, Y.~S.: 2012b,
\newblock {\em \apj} {\bf 753}, 148

\bibitem[Bovy et al.(2012c)]{bovy12c} 
Bovy, J., Rix, H.-W., Hogg, D.~W., et al.\ 2012c, \apj, 755, 115 

\bibitem[\protect\astroncite{{Brook} et~al.}{2012}]{brook12}
{Brook}, C.~B., {Stinson}, G.~S., {Gibson}, B., et al.: 2012,
\newblock {\em \mnras} {\bf 426}, 690

\bibitem[\protect\astroncite{{Brunetti} et~al.}{2011}]{brunetti11}
{Brunetti}, M., {Chiappini}, C., and {Pfenniger}, D.: 2011,
\newblock {\em \aap} {\bf 534}, A75

\bibitem[\protect\astroncite{{Carraro} et~al.}{1998}]{carraro98}
{Carraro}, G., {Ng}, Y.~K., and {Portinari}, L.: 1998,
\newblock {\em \mnras} {\bf 296}, 1045

\bibitem[\protect\astroncite{{Cavichia} et~al.}{2014}]{cavichia13}
{Cavichia}, O., {Moll{\'a}}, M., {Costa}, R.~D.~D., and {Maciel}, W.~J.: 2014,
\newblock {\em \mnras} {\bf 437}, 3688

\bibitem[\protect\astroncite{{Chen} et~al.}{2011}]{chen11}
{Chen}, Y.~Q., {Zhao}, G., {Carrell}, K., and {Zhao}, J.~K.: 2011,
\newblock {\em \aj} {\bf 142}, 184

\bibitem[\protect\astroncite{{Cheng} et~al.}{2012a}]{cheng12b}
{Cheng}, J.~Y., {Rockosi}, C.~M., {Morrison}, H.~L., et al.: 2012a,
\newblock {\em \apj} {\bf 752}, 51

\bibitem[\protect\astroncite{{Cheng} et~al.}{2012b}]{cheng12a}
{Cheng}, J.~Y., {Rockosi}, C.~M., {Morrison}, H.~L., et al.: 2012b,
\newblock {\em \apj} {\bf 746}, 149

\bibitem[\protect\astroncite{{Chiappini} et~al.}{1997}]{chiappini97}
{Chiappini}, C., {Matteucci}, F., and {Gratton}, R.: 1997,
\newblock {\em \apj} {\bf 477}, 765

\bibitem[\protect\astroncite{{Combes} et~al.}{1990}]{combes90}
{Combes}, F., {Debbasch}, F., {Friedli}, D., and {Pfenniger}, D.: 1990,
\newblock {\em \aap} {\bf 233}, 82

\bibitem[\protect\astroncite{{Daflon} and {Cunha}}{2004}]{daflon04}
{Daflon}, S. and {Cunha}, K.: 2004,
\newblock {\em \apj} {\bf 617}, 1115

\bibitem[\protect\astroncite{{de Jong} et~al.}{2012}]{dejong12}
{de Jong}, R.~S., {Bellido-Tirado}, O., and {Chiappini}, C. e.~a.: 2012,
\newblock 8446

\bibitem[\protect\astroncite{{Debattista} et~al.}{2006}]{debattista06}
{Debattista}, V.~P., {Mayer}, L., {Carollo}, C.~M., {Moore}, B., {Wadsley}, J.,
  and {Quinn}, T.: 2006,
\newblock {\em \apj} {\bf 645}, 209

\bibitem[\protect\astroncite{{Dehnen}}{2000}]{dehnen00}
{Dehnen}, W.: 2000,
\newblock {\em \aj} {\bf 119}, 800

\bibitem[\protect\astroncite{{Di Matteo} et~al.}{2013}]{dimatteo13}
{Di Matteo}, P., {Haywood}, M., {Combes}, F., {Semelin}, B., and {Snaith},
  O.~N.: 2013,
\newblock {\em \aap} {\bf 553}, A102

\bibitem[\protect\astroncite{{D'Onghia} et~al.}{2013}]{donghia13}
{D'Onghia}, E., {Vogelsberger}, M., and {Hernquist}, L.: 2013,
\newblock {\em \apj} {\bf 766}, 34

\bibitem[\protect\astroncite{{Foyle} et~al.}{2008}]{foyle08}
{Foyle}, K., {Courteau}, S., and {Thacker}, R.~J.: 2008,
\newblock {\em \mnras} {\bf 386}, 1821

\bibitem[\protect\astroncite{{Fran{\c c}ois} et~al.}{2004}]{francois04}
{Fran{\c c}ois}, P., {Matteucci}, F., {Cayrel}, R., {Spite}, M., {Spite}, F.,
  and {Chiappini}, C.: 2004,
\newblock {\em \aap} {\bf 421}, 613

\bibitem[\protect\astroncite{{Freeman} and {Bland-Hawthorn}}{2002}]{freeman02}
{Freeman}, K. and {Bland-Hawthorn}, J.: 2002,
\newblock {\em \araa} {\bf 40}, 487

\bibitem[\protect\astroncite{{Freeman} et~al.}{2010}]{freeman10}
{Freeman}, K., {Bland-Hawthorn}, J., and {Barden}, S.: 2010,
\newblock AAO Newsletter (February), in press

\bibitem[\protect\astroncite{{Frinchaboy} et~al.}{2013}]{frinchaboy13}
{Frinchaboy}, P.~M., {Thompson}, B., {Jackson}, K.~M., et al.: 2013,
\newblock {\em \apjl} {\bf 777}, L1

\bibitem[\protect\astroncite{{Fuhrmann}}{2011}]{fuhrmann11}
{Fuhrmann}, K.: 2011,
\newblock {\em \mnras} {\bf 414}, 2893

\bibitem[\protect\astroncite{{G{\'o}mez} et~al.}{2013}]{gomez13}
{G{\'o}mez}, F.~A., {Minchev}, I., {O'Shea}, B.~W., {Beers}, T.~C., {Bullock},
  J.~S., and {Purcell}, C.~W.: 2013,
\newblock {\em \mnras} {\bf 429}, 159

\bibitem[\protect\astroncite{{G{\'o}mez} et~al.}{2012a}]{gomez12b}
{G{\'o}mez}, F.~A., {Minchev}, I., {O'Shea}, B.~W., et al.: 2012a,
\newblock {\em \mnras} {\bf 423}, 3727

\bibitem[\protect\astroncite{{G{\'o}mez} et~al.}{2012b}]{gomez12a}
{G{\'o}mez}, F.~A., {Minchev}, I., {Villalobos}, {\'A}., {O'Shea}, B.~W., and
  {Williams}, M.~E.~K.: 2012b,
\newblock {\em \mnras} {\bf 419}, 2163

\bibitem[\protect\astroncite{{Hayden} et~al.}{2014}]{hayden14}
{Hayden}, M.~R., {Holtzman}, J.~A., {Bovy}, J., et al.: 2014,
\newblock {\em \aj} {\bf 147}, 116

\bibitem[\protect\astroncite{{Jacobson} et~al.}{2011}]{jacobson11}
{Jacobson}, H.~R., {Pilachowski}, C.~A., and {Friel}, E.~D.: 2011,
\newblock {\em \aj} {\bf 142}, 59

\bibitem[\protect\astroncite{{Kordopatis} et~al.}{2013}]{kordopatis13a}
{Kordopatis}, G., {Gilmore}, G., {Wyse}, R.~F.~G., {Steinmetz}, M., {Siebert},
  A., {Bienaym{\'e}}, O., {McMillan}, P.~J., {Minchev}, I., {Zwitter}, T.,
  {Gibson}, B.~K., {Seabroke}, G., {Grebel}, E.~K., {Bland-Hawthorn}, J.,
  {Boeche}, C., {Freeman}, K.~C., {Munari}, U., {Navarro}, J.~F., {Parker}, Q.,
  {Reid}, W.~A., and {Siviero}, A.: 2013,
\newblock {\em \mnras} {\bf 436}, 3231

\bibitem[\protect\astroncite{{Kordopatis} et~al.}{2011}]{kordopatis11}
{Kordopatis}, G., {Recio-Blanco}, A., {de Laverny}, P., {Gilmore}, G., {Hill},
  V., {Wyse}, R.~F.~G., {Helmi}, A., {Bijaoui}, A., {Zoccali}, M., and
  {Bienaym{\'e}}, O.: 2011,
\newblock {\em \aap} {\bf 535}, A107

\bibitem[\protect\astroncite{{Kubryk} et~al.}{2013}]{kubryk13}
{Kubryk}, M., {Prantzos}, N., and {Athanassoula}, E.: 2013,
\newblock {\em \mnras} {\bf 436}, 1479

\bibitem[\protect\astroncite{{Lee} et~al.}{2011}]{lee11}
{Lee}, Y.~S., {Beers}, T.~C., {Allende Prieto}, C., {Lai}, D.~K., {Rockosi},
  C.~M., {Morrison}, H.~L., {Johnson}, J.~A., {An}, D., {Sivarani}, T., and
  {Yanny}, B.: 2011,
\newblock {\em \aj} {\bf 141}, 90

\bibitem[\protect\astroncite{{Lemasle} et~al.}{2013}]{lemasle13}
{Lemasle}, B., {Fran{\c c}ois}, P., {Genovali}, K., et al.: 2013,
\newblock {\em \aap} {\bf 558}, A31

\bibitem[\protect\astroncite{{Liu} and {van de Ven}}{2012}]{liu12}
{Liu}, C. and {van de Ven}, G.: 2012,
\newblock {\em \mnras} {\bf 425}, 2144

\bibitem[\protect\astroncite{{Luck} and {Lambert}}{2011}]{luck11}
{Luck}, R.~E. and {Lambert}, D.~L.: 2011,
\newblock {\em \aj} {\bf 142}, 136

\bibitem[\protect\astroncite{{Martig} et~al.}{2012}]{martig12}
{Martig}, M., {Bournaud}, F., {Croton}, D.~J., {Dekel}, A., and {Teyssier}, R.:
  2012,
\newblock {\em \apj} {\bf 756}, 26

\bibitem[\protect\astroncite{{Martig} et~al.}{2009}]{martig09}
{Martig}, M., {Bournaud}, F., {Teyssier}, R., and {Dekel}, A.: 2009,
\newblock {\em \apj} {\bf 707}, 250

\bibitem[\protect\astroncite{{Martig} et~al.}{2014a}]{martig14a}
{Martig}, M., {Minchev}, I., and {Flynn}, C.: 2014a,
\newblock {\em \mnras} {\bf 442}, 2474

\bibitem[\protect\astroncite{{Martig} et~al.}{2014b}]{martig14b}
{Martig}, M., {Minchev}, I., and {Flynn}, C.: 2014b,
\newblock {\em arXiv: 1405.1727}

\bibitem[\protect\astroncite{{Masset} and {Tagger}}{1997}]{masset97}
{Masset}, F. and {Tagger}, M.: 1997,
\newblock {\em \aap} {\bf 322}, 442

\bibitem[\protect\astroncite{{Matteucci}}{2012}]{matteucci12}
{Matteucci}, F.: 2012,
\newblock {\em {Chemical Evolution of Galaxies}}

\bibitem[\protect\astroncite{{Minchev} et~al.}{2010}]{minchev10}
{Minchev}, I., {Boily}, C., {Siebert}, A., and {Bienayme}, O.: 2010,
\newblock {\em \mnras} {\bf 407}, 2122

\bibitem[\protect\astroncite{{Minchev} et~al.}{2013}]{mcm13}
{Minchev}, I., {Chiappini}, C., and {Martig}, M.: 2013,
\newblock {\em \aap} {\bf 558}, A9

\bibitem[\protect\astroncite{{Minchev} et~al.}{2014}]{minchev14a}
{Minchev}, I., {Chiappini}, C., {Martig}, M., {Steinmetz}, M., {de Jong},
  R.~S., {Boeche}, C., {Scannapieco}, C., {Zwitter}, T., {Wyse}, R.~F.~G.,
  {Binney}, J.~J., {Bland-Hawthorn}, J., {Bienayme}, O., {Famaey}, B.,
  {Gibson}, B.~K., {Grebel}, E.~K., {Gilmore}, G., {Helmi}, A., {Kordopatis},
  G., {Lee}, Y.~S., {Munari}, U., {Navarro}, J.~F., {Parker}, Q.~A., {Quillen},
  A.~C., {Reid}, W.~A., {Siebert}, A., {Siviero}, A., {Seabroke}, G., {Watson},
  F., and {Williams}, M.: 2014,
\newblock {\em \apjl} {\bf 781}, L20

\bibitem[\protect\astroncite{{Minchev} and {Famaey}}{2010}]{mf10}
{Minchev}, I. and {Famaey}, B.: 2010,
\newblock {\em \apj} {\bf 722}, 112

\bibitem[\protect\astroncite{{Minchev} et~al.}{2011}]{minchev11a}
{Minchev}, I., {Famaey}, B., {Combes}, F., {Di Matteo}, P., {Mouhcine}, M., and
  {Wozniak}, H.: 2011,
\newblock {\em \aap} {\bf 527}, 147

\bibitem[\protect\astroncite{{Minchev} et~al.}{2012a}]{minchev12b}
{Minchev}, I., {Famaey}, B., {Quillen}, A.~C., {Dehnen}, W., {Martig}, M., and
  {Siebert}, A.: 2012a,
\newblock {\em \aap} {\bf 548}, A127

\bibitem[\protect\astroncite{{Minchev} et~al.}{2012b}]{minchev12a}
{Minchev}, I., {Famaey}, B., {Quillen}, A.~C., {Di Matteo}, P., {Combes}, F.,
  {Vlaji{\'c}}, M., {Erwin}, P., and {Bland-Hawthorn}, J.: 2012b,
\newblock {\em \aap} {\bf 548}, A126

\bibitem[\protect\astroncite{{Minchev} et~al.}{2007}]{mnq07}
{Minchev}, I., {Nordhaus}, J., and {Quillen}, A.~C.: 2007,
\newblock {\em \apjl} {\bf 664}, L31

\bibitem[\protect\astroncite{{Minchev} et~al.}{2009}]{minchev09}
{Minchev}, I., {Quillen}, A.~C., {Williams}, M., {Freeman}, K.~C., {Nordhaus},
  J., {Siebert}, A., and {Bienaym{\'e}}, O.: 2009,
\newblock {\em \mnras} {\bf 396}, L56

\bibitem[\protect\astroncite{{Pedicelli} et~al.}{2009}]{pedicelli09}
{Pedicelli}, S., {Bono}, G., {Lemasle}, B., et al.: 2009,
\newblock {\em \aap} {\bf 504}, 81

\bibitem[\protect\astroncite{{Perryman} et~al.}{2001}]{perryman01}
{Perryman}, M.~A.~C., {de Boer}, K.~S., {Gilmore}, G., et al.: 2001,
\newblock {\em \aap} {\bf 369}, 339

\bibitem[\protect\astroncite{{Quillen} et~al.}{2011}]{quillen11}
{Quillen}, A.~C., {Dougherty}, J., {Bagley}, M.~B., {Minchev}, I., and
  {Comparetta}, J.: 2011,
\newblock {\em \mnras} {\bf 417}, 762

\bibitem[\protect\astroncite{{Quillen} et~al.}{2014}]{quillen14}
{Quillen}, A.~C., {Minchev}, I., {Sharma}, S., {Qin}, Y.-J., and {Di Matteo},
  P.: 2014,
\newblock {\em \mnras} {\bf 437}, 1284

\bibitem[\protect\astroncite{{Ramya} et~al.}{2012}]{ramya12}
{Ramya}, P., {Reddy}, B.~E., and {Lambert}, D.~L.: 2012,
\newblock {\em \mnras} {\bf 425}, 3188

\bibitem[\protect\astroncite{{Recio-Blanco} et~al.}{2014}]{recio-blanco14}
{Recio-Blanco}, A., {de Laverny}, P., {Kordopatis}, G., et al.: 2014,
\newblock {\em \aap} {\bf 567}, A5

\bibitem[\protect\astroncite{{Rix} and {Bovy}}{2013}]{rix13}
{Rix}, H.-W. and {Bovy}, J.: 2013,
\newblock {\em \aapr} {\bf 21}, 61

\bibitem[\protect\astroncite{{Ro{\v s}kar} et~al.}{2008}]{roskar08a}
{Ro{\v s}kar}, R., {Debattista}, V.~P., {Quinn}, T.~R., {Stinson}, G.~S., and
  {Wadsley}, J.: 2008,
\newblock {\em \apjl} {\bf 684}, L79

\bibitem[\protect\astroncite{{Sellwood} and {Binney}}{2002}]{sellwood02}
{Sellwood}, J.~A. and {Binney}, J.~J.: 2002,
\newblock {\em \mnras} {\bf 336}, 785

\bibitem[\protect\astroncite{{Stasi{\'n}ska} et~al.}{2012}]{stasinska12}
{Stasi{\'n}ska}, G., {Prantzos}, N., {Meynet}, G., et al.: 2012,
\newblock in {\em EAS Publications Series}, Vol.~54 of {\em EAS Publications
  Series}, pp 255--317

\bibitem[\protect\astroncite{{Steinmetz}}{2012}]{steinmetz12}
{Steinmetz}, M.: 2012,
\newblock {\em Astronomische Nachrichten} {\bf 333}, 523

\bibitem[\protect\astroncite{{Stinson} et~al.}{2013}]{stinson13}
{Stinson}, G.~S., {Bovy}, J., {Rix}, H.-W., et al.: 2013,
\newblock {\em \mnras}

\bibitem[\protect\astroncite{{Vera-Ciro} et~al.}{2014}]{vera-ciro14}
{Vera-Ciro}, C., {D'Onghia}, E., {Navarro}, J., and {Abadi}, M.: 2014,
\newblock {\em ArXiv: 1405.3317}

\bibitem[\protect\astroncite{{Widrow} et~al.}{2012}]{widrow12}
{Widrow}, L.~M., {Gardner}, S., {Yanny}, B., {Dodelson}, S., and {Chen}, H.-Y.:
  2012,
\newblock {\em \apjl} {\bf 750}, L41

\bibitem[\protect\astroncite{{Yanny} and {Rockosi}}{2009}]{yanny09}
{Yanny}, B. and {Rockosi}, C., N.~H.~J. e.~a.: 2009,
\newblock {\em \aj} {\bf 137}, 4377

\bibitem[\protect\astroncite{{Yong} et~al.}{2012}]{yong12}
{Yong}, D., {Carney}, B.~W., and {Friel}, E.~D.: 2012,
\newblock {\em \aj} {\bf 144}, 95

\end{thebibliography}

\begin{appendix}

\section{Comparison between the chemical model and simulation SFHs}

\begin{figure}
\includegraphics[width=8cm]{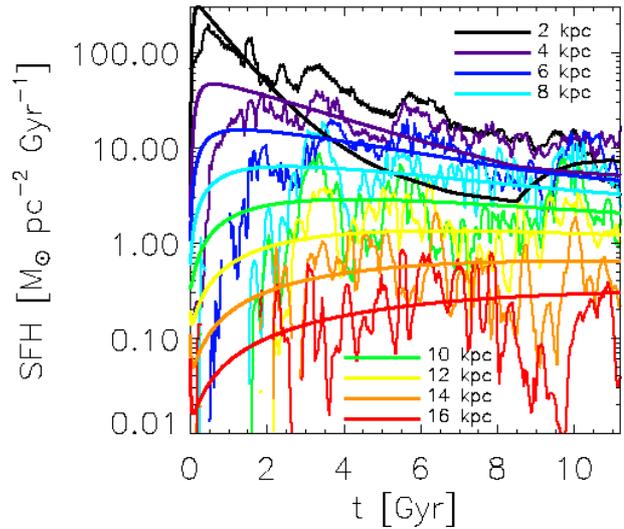}
\caption{
Comparison between the SFH (as a function of cosmic time) of the simulation and the chemical evolution model. Different colors correspond to different disk radii, as indicated. Good overall agreement is found for most radii.    
}
\label{fig:sfh}      
\end{figure}

In Fig.~\ref{fig:sfh} we compare the SFH (as a function of cosmic time) of the simulation with that used for the chemical evolution model. Different colors correspond to different disk radii, as indicated. Good overall agreement is found for most radii. Some discrepancy can be seen at 2 and 4 kpc approximately at $t>3$ and $t>5$~Gyr, respectively, where the simulation SFH is somewhat stronger. On the other hand, at early times the simulation SFH for $r\ge4$~kpc is lower than the chemical model SFH. This means that the disk inside-out growth in somewhat delayed in the simulation. As discussed in paper I, we have assumed that these differences between the simulation and chemical model are not crucial for the dynamics. However, we weight the stars in the simulation to satisfy the SFH in the chemical model, which is important for obtaining the correct mixture of populations corresponding to the chemistry.

\section{Variations in chemo-kinematic properties with radius and distance from the disk midplane}
\label{sec:a1}

\begin{figure*}
\includegraphics[width=18cm]{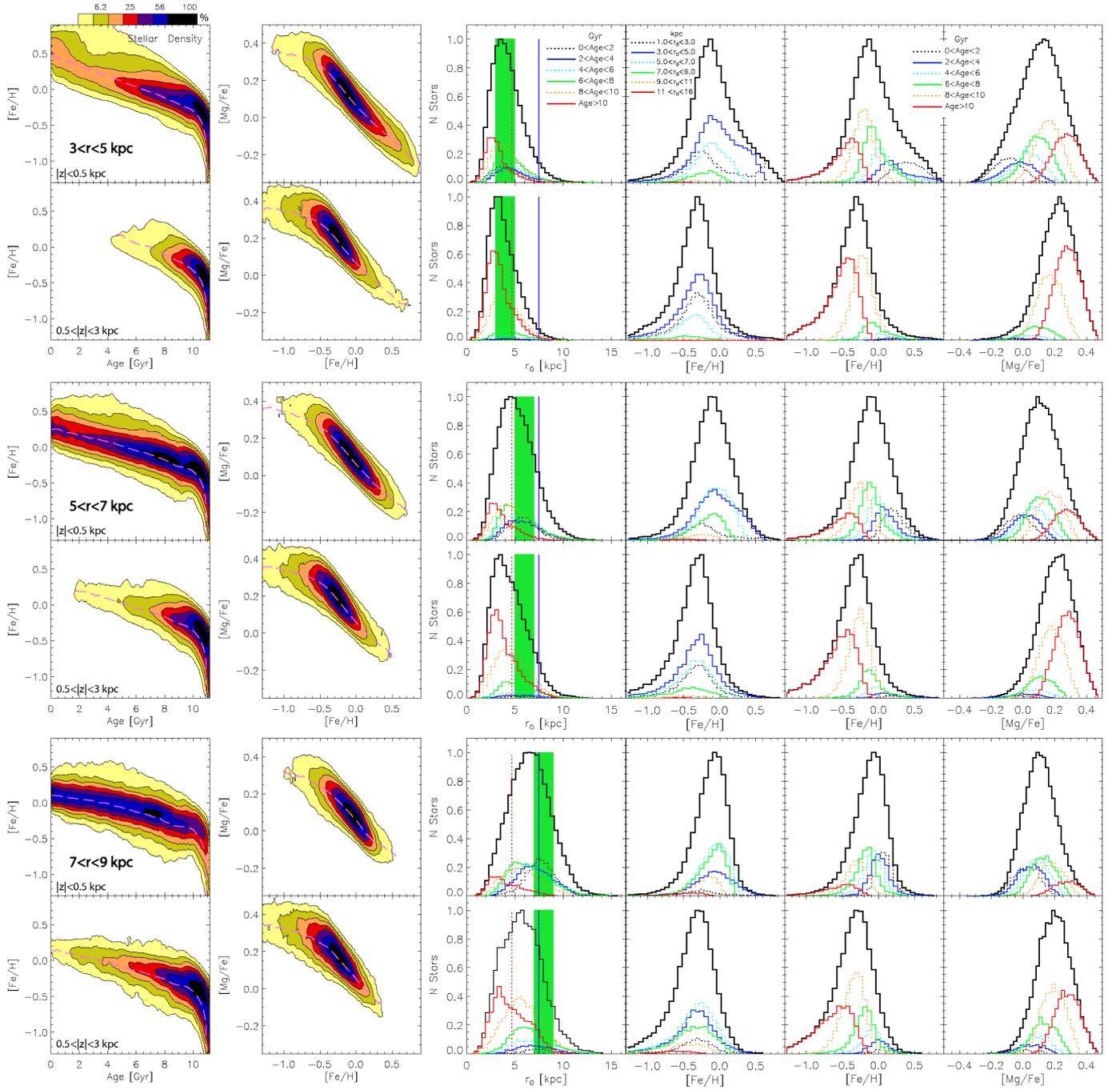}
\caption{
Predictions for chemo-kinematic properties at different distances from the Galactic center (see also Fig.~\ref{fig:chemz2}). For each 2-kpc annulus a sample close to the plane ($|z|<0.5$~kpc) and farther from the plane ($0.5<|z|<3$~kpc) is plotted. The dashed-pink lines in the first two columns show the mean. 
}
\label{fig:chemz1}      
\end{figure*}

\begin{figure*}
\includegraphics[width=18cm]{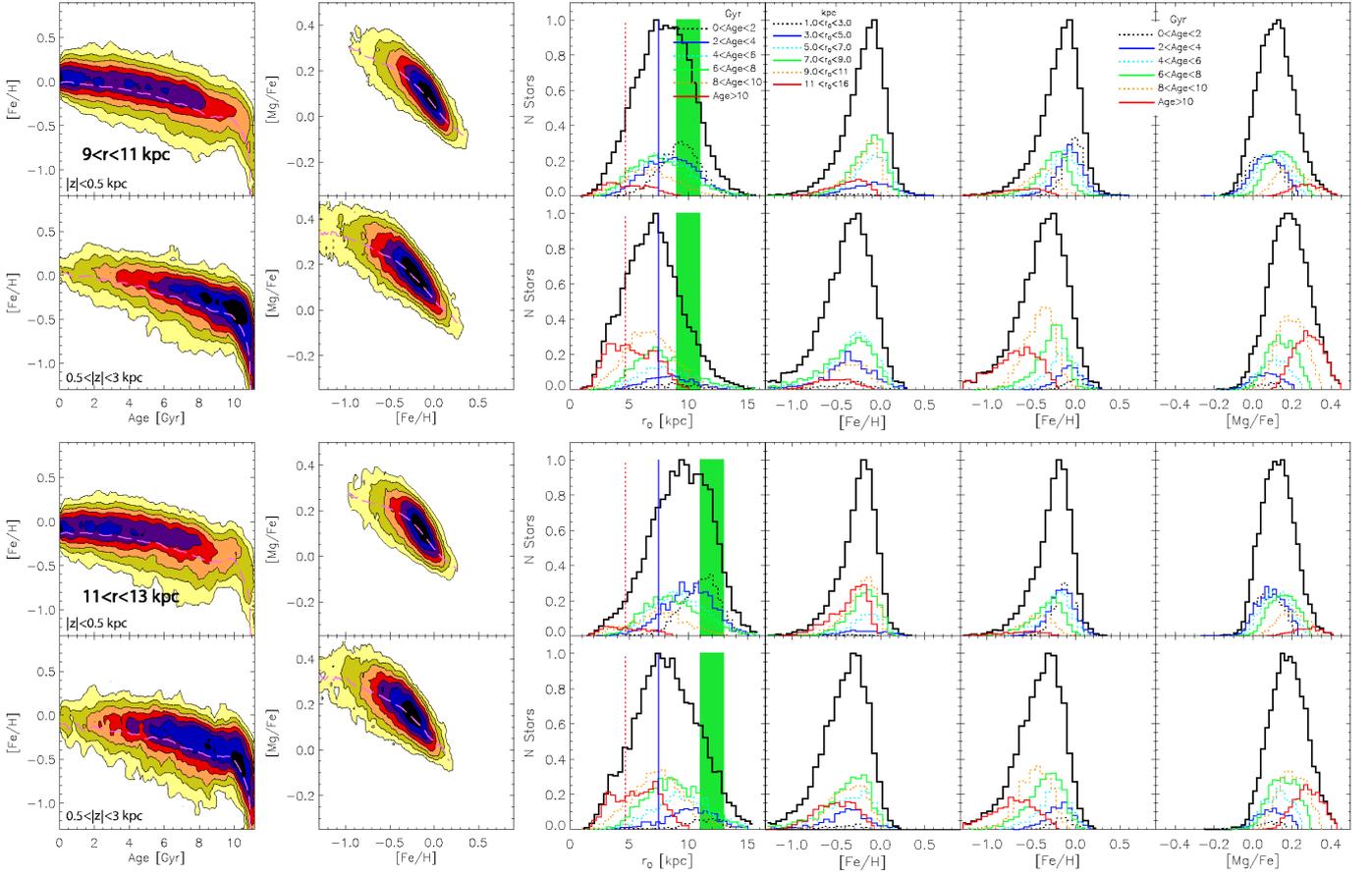}
\caption{
As Fig.~\ref{fig:chemz1} but for samples with $9<r<11$ and $11<r<13$~kpc.
}
\label{fig:chemz2}      
\end{figure*}

Figures~\ref{fig:chemz1} and \ref{fig:chemz2} present similar information to Figures~\ref{fig:chem} and \ref{fig:chem2}, but for two different distances from the disk midplane, as indicated. For each 2-kpc annulus a sample close to the plane ($|z|<0.5$~kpc) and farther from the plane ($0.5<|z|<3$~kpc) is plotted. The dashed-pink lines in the first two columns show the mean. The decrease in age with Galactic radius can be seen in the densest contour levels (especially at $0.5<|z|<3$~kpc), which do not extend to young ages for the inner radial bins. For the samples away from the plane shifts in the distributions are seen to lower birth-radii, lower metallicities and higher [Mg/Fe], as expected for older populations.  

We note that the [Fe/H]-[Mg/Fe] plane does not show a discontinuity, as frequently seen in observations. We demonstrated in paper I that a kinematically biased samples can give rise to this division between the thin and thick disks in the solar vicinity. However, as currently debated in the literature \citep{fuhrmann11, anders14, bovy12b}, this gap may not be an artifact, but a real feature reflecting variations in the SFH of the MW disk. If this is confirmed in observations, it will serve as an additional constraint for our model, requiring the modification of its SFH and early chemistry. Except for the chemical discontinuity, no significant changes to the results presented in paper I and the current work are expected.

\end{appendix}

\end{document}